\documentclass[trackchanges, twocolumn, twocolappendix]{aastex701}
\usepackage{amsmath}

\defcitealias{Matthews_2021b}{M21b}
\defcitealias{2014ARA&A..52..415M}{MD}
\begin{document}

\title{The Radio-FIR Correlation in the Context of Deep Radio Source Counts}

\author[orcid=0000-0002-6369-6266,sname='North America']{Tirth D. Surti}
\affiliation{Cahill Center for Astronomy and Astrophysics, California Institute of Technology, 1216 E California Blvd, Pasadena, CA 91125,
USA}
\email[show]{tsurti@caltech.edu}  

\author[orcid=0000-0002-7252-5485]{Vikram Ravi} 
\affiliation{Cahill Center for Astronomy and Astrophysics, California Institute of Technology, 1216 E California Blvd, Pasadena, CA 91125,
USA}
\email{vikram@caltech.edu}

\author[orcid=0000-0002-7252-5485]{Allison Matthews} 
\affiliation{Carnegie Observatories, 813 Santa Barbara Street, Pasadena, CA 91101, USA}
\email{amatthews@carnegiescience.edu}

\author[orcid=0000-0001-8596-1756]{Viviana A. Rosero}
\affiliation{Cahill Center for Astronomy and Astrophysics, California Institute of Technology, 1216 E California Blvd, Pasadena, CA 91125,
USA}
\email{vrosero@caltech.edu}

\begin{abstract}
Increasingly deep, confusion-limited radio surveys have pushed direct radio source-count measurements down to tens of $\mu$Jy at 1.4 GHz. Confusion-noise P(D) analyses extend the statistical counts to below $1\,\mathrm{\mu Jy}$. Radio source counts have allowed for constraints on the radio-derived star formation rate density (SFRD) history through models of the backwards evolution of the local radio luminosity function, using the radio-FIR correlation, $q \propto  \log(L_{\mathrm{FIR}}/L_{1.4})$, to convert radio luminosities to FIR luminosities and hence star-formation rates. Recent deep radio source counts from \textit{MeerKAT} suggest a potential tension in the SFRD history between radio and UV/IR measurements at $1\lesssim z\lesssim 2$. This corresponds to a $\gtrsim 3\sigma$ discrepancy between the predicted and measured source counts near the star forming galaxy source count $S^2n(S)$ peak of ${\sim}30\,\mathrm{\mu Jy}$ under both a pure luminosity (PLE) and combined luminosity and density evolution (LADE). We consider what the requirement of agreement between radio source counts and the observed UV/IR SFRD indicates about the redshift evolution of the radio-FIR correlation and its intrinsic scatter out to $z=3$. We introduce a radio-luminosity based parameterization to $q_{\mathrm{FIR}}(z)$ based on changing thermal radio fractions alone that agrees with the observed stellar-mass dependent $q_{\mathrm{FIR}}(z)$ better than a nonevolving or decreasing $q_{\mathrm{FIR}}(z)$. Despite this, we find that a decreasing $q_{\mathrm{FIR}}(z)$ at fixed radio luminosity provides better agreement between source counts and the observed SFRD, while a $q_{\mathrm{FIR}}(z)$ that breaks down due to cosmic-ray losses requires an intrinsic scatter up to $\sigma_q\approx 0.3\,\mathrm{dex}$.

%We introduce a purely radio-luminosity based parameterization of the redshift evolution of the radio-FIR correlation based on changing cosmic-ray losses. We find evidence (${\gtrsim}2\sigma$) that an evolution in the radio-FIR correlation consistent with a mild decrease in$q_{\mathrm{FIR}}$out to $z{\sim}2$ arising from strengthening magnetic fields can mitigate the source count tension. We additionally show that intrinsic scatter in the radio-FIR correlation is likely bounded $\sigma_q\lesssim 0.3\,\mathrm{dex}$ at these redshifts if$q_{\mathrm{FIR}}$decreases. Although we find no evidence that current radio source counts imply a breakdown in the radio-FIR correlation, future deep radio surveys from the Deep Synoptic Array (DSA) will be able to push radio source counts down to several nJy, providing stronger constraints on the allowed evolution.
\end{abstract}



\section{Introduction}
Differential radio source counts down to micro-Jansky fluxes provide a fundamental constraint on the evolution of star formation rate density (SFRD), because they indicate the integrated sum of the flux density contributions from star forming galaxies (SFGs) across all redshifts. Deep radio surveys \citep[e.g.,][]{Condon_2012,10.1093/mnras/stu470, 2017A&A...602A...1S, lofar_counts, Matthews_2021a, 10.1093/mnras/stae2528, 2026A&A...707A.198S}, further aided by a P(D) analysis of confusion-limited images, have measured differential source count contributions $S^2n(S)$ over flux densities dominated by both active galactic nuclei (AGN) and SFGs at ${\sim}100\,\mathrm{mJy}$ and ${\sim}30\,\mathrm{\mu Jy}$ at 1.4\, GHz, respectively. SFRD measurements constrained by multiple UV and infrared (IR) surveys indicate an SFRD that rose sharply and peaked near $z{\sim}2$ before decaying with $e$-folding timescale $\tau{\sim}4$ Gyr to a local SFRD $\psi_0{\sim}0.015\,\mathrm{M_{\odot}\,yr^{-1}\,Mpc^{-3}}$ \citep[][hereafter MD]{2014ARA&A..52..415M} under a fixed Salpeter initial mass function \citep[IMF,][]{1955ApJ...121..161S}. This compilation is a standard for comparison for radio-derived cosmic SFRDs that have become possible with deeper and nearly photometrically complete radio surveys out to higher redshifts \citep[e.g.,][]{novak_2017, 10.1093/mnras/stad1602, 2026MNRAS.548ag616T}.

Given the immunity of radio emission to dust extinction and contamination by older stellar populations, the likely origin of radio emission from HII regions (thermal free--free) and accelerated cosmic-rays from supernovae \citep[nonthermal synchrotron,][]{1975ApJ...200L.127H} suggests that radio emission can also trace star formation \citep{1992ARA&A..30..575C}. This is most conveniently parameterized in terms of the observed tight radio--far infrared (FIR) correlation defined by the parameter
\begin{equation}
q_{\mathrm{FIR}} = \log\left(\frac{L_{\mathrm{FIR}}/(3.75\times 10^{12}\,\mathrm{Hz})}{L_{1.4}}\right),
\end{equation}
which is used to first convert to an equivalent FIR luminosity and then a SFR using a known $q_{\mathrm{FIR}}$. This correlation holds for both FIR luminosities \citep[e.g., calculated using a combination of 60 and 100\,$\mathrm{\mu m}$ IRAS flux densities,][]{1989cgqo.book.....F} and total-infrared (TIR) luminosities integrated from 8-1000$\,\mathrm{\mu m}$, as they are observed to have a typical ratio $\langle L_{\mathrm{TIR}}/L_{\mathrm{FIR}}\rangle\approx 2$ \citep{2003ApJ...586..794B, 10.1093/mnras/staf1536}.

The radio-FIR correlation was originally observed using local star forming galaxies \citep{1973A&A....29..263V, 1984AJ.....89.1520R}, showing a remarkably small dispersion of ${\sim}38\%$ \citep{1985ApJ...298L...7H}. Under the assumption of a fixed 1.4\,GHz spectral index, $q_{\mathrm{FIR}}$ is observed to exhibit a mild redshift evolution decreasing out to $z{\sim}3$ and potentially beyond \citep[e.g.,][]{2017A&A...602A...4D, 2017MNRAS.469.3468C}. More fundamental is likely its relation to SFG mass (radio luminosity), with $q_{\mathrm{FIR}}$ observed to rise with smaller masses (radio luminosities) \citep{2021AA...647A.123D} and potentially flatten above $\log(L_{1.4}/(\mathrm{W\, Hz^{-1}}))\geq22.5$ \citep{Matthews_2021b, 10.1093/mnras/staf1536}. At fixed stellar mass, \cite{2021AA...647A.123D} show that any redshift evolution in $q_{\mathrm{FIR}}(z)$ is significantly weaker due to the selection effect of probing larger stellar masses at higher redshifts.

While the radio-FIR correlation is based on the total rest-frame 1.4 GHz radio luminosities which is dominated by nonthermal synchrotron emission in normal SFGs, it can also be interpreted in terms of the fraction of radio emission that is thermal at 1.4 GHz. Given the relationship between ionizing photon rate from young stars in HII regions and thermal radio emission and separately to SFR on the ${\lesssim100}\,\mathrm{Myr}$ timescale traced by IR emission, $L_{1.4}^{T}\propto L_{\mathrm{TIR}}$ \citep{2009ApJ...706..482M, 2024ApJ...975...15Y}. Consequently, assuming $\langle L_{\mathrm{TIR}}/L_{\mathrm{FIR}}\rangle\approx 2$, the $q_{\mathrm{FIR}}$ parameter can be related to the thermal fraction as $q\approx q_0+\log(f_t)$ with 
\begin{equation}
q_0 = \log\left(\frac{L_{\mathrm{FIR}}/(3.75\times10^{12}\,\mathrm{Hz})}{L_{1.4}^{T}}\right)\approx 3.32,
\end{equation}
which reproduces the typical observed $\langle q_{\mathrm{FIR}}\rangle \approx 2.34$ for the average $\langle f_t\rangle \approx 10\%$ thermal fraction at 1.4 GHz consistent with that observed in local SFGs \citep{1990ApJ...357...97C}. This naturally frames any potential redshift evolution in $q_{\mathrm{FIR}}(z)$ in terms of a redshift evolution in the relative FIR-to-thermal radio calibration $q_0$ and in the thermal fractions $f_t$, the latter of which is likely driven by redshift-dependent cosmic-ray losses. Depending on the SFG sample, the scatter in the local radio-luminosity dependent radio-FIR correlation is inferred to be as low as ${\sim}0.13$-$0.16\,\mathrm{dex}$ in individual radio luminosity bins \citep{2021MNRAS.504..118M}. However, \cite{2021AA...647A.123D} infer a slightly larger scatter ${\sim} 0.21$-$0.22\,\mathrm{dex}$ about the mean $q_{\mathrm{FIR}}(z)$ redshift evolution at fixed stellar mass.

%\textcolor{red}{Deviations from the thermal radio-FIR calibration $q_0\approx 3.32$ are expected but not well-constrained with redshift and even locally. $q_0$ is expected to decrease in less luminous (massive) SFGs as the reduced dust content makes FIR emission underestimate star formation rates relative to thermal radio emission \citep[e.g., as inferred by the smaller IR-to-UV ratio, IRX, in less luminous and massive SFGs][]{2003ApJ...586..794B, 2020ApJ...902..112B}. In more luminous and dusty SFGs, increased dust absorption of ionizing photons that would otherwise contribute to ionizing neutral hydrogen may increase $q_0$ \citep[e.g.,][]{Inoue_2001, 2003ApJ...583..727D}, corresponding to a reduced thermal radio fraction relative to the observed $q_{\mathrm{FIR}}$. Alternatively, there may be increased contributions to dust heating by old stellar populations, but at least locally, this appears to have little effect on the radio-FIR correlation \citep[e.g.,][]{2003ApJ...586..794B}. At higher radio luminosities, SFGs may become increasingly contaminated by AGN \citep{Dey_2024}. Classifications of SFGs and AGN based on $q_{\mathrm{FIR}}$ appear to be successful at preventing radio-loud AGN in contaminating SFG samples \citep{2026A&A...706A.111P}, but it is not clear whether such AGN could still contribute an FIR excess.}

As previously done for infrared--based source counts \citep[e.g.,][]{2001MNRAS.325.1511P, ir_back2}, radio source counts can be modeled by evolving the local 1.4\,GHz radio luminosity function (LF) backwards under an assumed pure luminosity and density evolution such that the radio LF at any redshift $z$ can be expressed in terms of the local LF \citep{novak_2017, Matthews_2021b, 2026MNRAS.548ag616T}. For radio LFs, because $L_{\star}$ SFGs fall out of survey detection limits above $z{\sim}1$, the luminosity and density evolution tend to be strongly degenerate. Consequently, some analyses suggest that a pure luminosity evolution (PLE) is sufficient to explain radio LFs constructed out to $z{\sim}4$ \citep{novak_2017, 2022ApJ...941...10V, 2026MNRAS.548ag616T}, while other analyses suggest that the density starts to drop rapidly at $z{\sim}1$ \citep[e.g.,][]{2024A&A...683A.174W}, requiring a luminosity and density evolution (LADE).

In this paper, we build on the work of \cite{Matthews_2021b}, hereafter M21b, which finds a radio--derived SFR history that may be in tension with UV/IR measurements but does not consider a potentially evolving SFR calibration due to an evolving radio-FIR correlation nor the effects of intrinsic scatter in the correlation. In Section 2, we briefly summarize the radio source count model and in Section 3, we use this model to show the discrepancy between the radio source count model and the UV/IR SFRD history under a nonevolving radio-FIR correlation and both a PLE and LADE of the local radio LF. We use this discrepancy as an opportunity to consider the possibility of a redshift-evolving radio-FIR correlation on the agreement with radio source counts, whose models are presented in Section 4. In Section 5, we consider the effect of intrinsic scatter in the radio-FIR correlation when converting between radio and IR LFs. In Section 6, we discuss the impact of these effects on the agreement with radio source counts, and in Section 7 we explicitly compare the different models and evaluate their assumptions.

To remain consistent with \citetalias{Matthews_2021b} and \citetalias{2014ARA&A..52..415M}, we assume a flat $\Lambda$CDM cosmology with $H_0=70$ and $\Omega_{m,0}=0.3$. We take the spectral index $\alpha$ to be such that $S_{\nu}{\sim\nu^{\alpha}},$ so $\alpha<0$ for optically thin thermal free-free and nonthermal synchrotron emission in SFGs. We also assume the Salpeter IMF \citep{1955ApJ...121..161S} in all SFRD calculations for consistency with \citetalias{2014ARA&A..52..415M}.

\section{Model Overview}
We aim to model the $S^2n(S)$ radio source counts derived from both the MeerKAT DEEP2 survey \citep{2020ApJ...888...61M} and the NRAO Very Large Array (VLA) Sky Survey \citep[NVSS,][]{1998AJ....115.1693C} catalogs \citep[radio source count data can be found in Tables 5 and 6 of][]{Matthews_2021a}. The MeerKAT DEEP2 field  was imaged at a central frequency of 1.27 GHz over a primary beam FWHM of ${\sim}69^{\prime}$ and angular resolution of $\theta_{\mathrm{FWHM}}=7.6^{\prime\prime}$. Achieving a central thermal noise of $\sigma_n\approx 0.56\pm 0.01\,\mathrm{\mu Jy\, beam^{-1}}$, the survey is confusion limited, so direct source counts are only available down to $S_{1.4}\approx 10\,\mathrm{\mu Jy}$ \citep{Matthews_2021a}. NVSS supplies direct source counts for flux densities $\gtrsim 3\,\mathrm{mJy}$, close to its nominal $5\sigma$ detection threshold. A confusion P(D)-based extrapolation on the DEEP2 image enables additional constraints on the source counts down to ${\sim}0.25\,\mathrm{\mu Jy}$, well below the confusion limit. Uncertainties on the direct source counts arise from both Poisson and flux density calibration uncertainties. Uncertainties on the source counts extrapolated from the P(D) were determined by varying the thermal noise and finding the set of radio source count models $S^2n(S)$ below the confusion limit consistent with the observed P(D).

We adopt the procedure of \citetalias{Matthews_2021b}, whereby the local radio LF for SFGs and AGN are both evolved backwards in time under a luminosity and density evolution. This evolution is different for SFGs and AGN. We refer the reader to \citetalias{Matthews_2021b} for the full description of the radio source count model, but we summarize the relevant parts of the model briefly in this section and modifications made to the original procedure.

\subsection{Radio Luminosity Function Model}
\begin{figure}
    \centering
    \includegraphics[width=\linewidth]{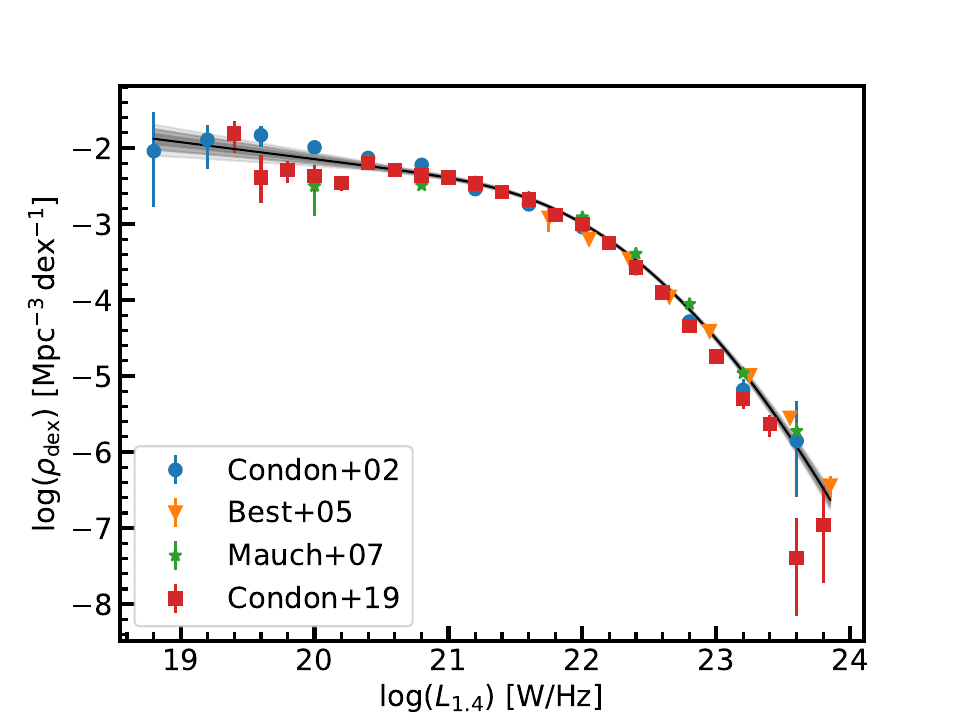}
    \caption{Combined fit to the local SFG radio LF using independent source counts provided by \cite{2002AJ....124..675C, 2005MNRAS.362....9B, 2007MNRAS.375..931M, 2019ApJ...872..148C}, spanning nearly $5\,\mathrm{dex}$ in radio luminosities. We have overlaid the 1, 2, and 3$\sigma$ contours in decreasing opacity.}
    \label{fig:fig1}
\end{figure}

If $\rho_{\mathrm{dex}}(L_{\nu}|z)$ is the radio LF at redshift $z$, then the luminosity density function is $u_{\mathrm{dex}}(L_{\nu}|z)=L_{\nu}\rho_{\mathrm{dex}}(L_{\nu}|z)$, which has dimensions of spectral luminosity per volume per dex of radio luminosity. Assuming a pure luminosity evolution $f(z)$ and density evolution $g(z)$, the luminosity density function at any redshift can be related to the local luminosity density function as
\begin{equation}
    u_{\mathrm{dex}}(L_{\nu}|z) = f(z)g(z)u_{\mathrm{dex}}(L_{\nu}/f(z) | 0).
\end{equation}
For the local AGN LF, we assume the parameters used by M21b, originally derived from \cite{2019ApJ...872..148C}. For SFGs, to reduce the impact of cosmic variance, we combine independent 1.4\,GHz radio LFs derived from \cite{2002AJ....124..675C, 2005MNRAS.362....9B, 2007MNRAS.375..931M, 2019ApJ...872..148C} following the approach of \cite{novak_2017} and fit the standard functional form
\begin{equation}
    \rho_{\mathrm{dex}}(L_{\nu}|0) = \rho_{\star}\left(\frac{L_{\nu}}{L_{\star}}\right)^{1-\alpha}\exp\left(-\frac{1}{2\sigma^2}\log^2\left(1 + \frac{L_{\nu}}{L_{\star}}\right)\right)
\end{equation}
originally from \cite{1990MNRAS.242..318S}. Here, $\rho_{\star}$ is the density at the cut-off luminosity $L_{\star}$, $\alpha$ controls the slope below $L_{\star}$, and $\sigma$ controls the width or slope above $L_{\star}$. A Markov Chain Monte Carlo (MCMC) fit using \textsc{emcee} \citep{2013PASP..125..306F} with symmetrized errors and a Gaussian log-likelihood gives $\log\rho_{\star} = -2.41\pm0.03$ ($\mathrm{Mpc^{-3}\, dex^{-1}}$), $\log L_{\star}=21.16\pm0.06$ ($\mathrm{W\,Hz^{-1}}$), $\alpha=1.22\pm 0.03$, and $\sigma=0.657\pm0.012$. The high-luminosity slope $\sigma$ is higher by ${\sim}0.1$ compared with  \citet{2019ApJ...872..148C}, resulting in a minimal ${\sim}0.07\,\mathrm{dex}$ increase in the local SFRD for a given IMF. Hereafter, we keep these parameters fixed to reduce the parameter space complexity when fitting for the luminosity and density evolution. In Figure \ref{fig:fig1}, we show the confidence intervals of the posterior plotted over the combined dataset, which spans ${\sim}5$ decades in luminosity.

\subsection{Radio Source Count Model}
Under a spectral index distribution $p(\alpha|z)$ and a luminosity density function $u_{\mathrm{dex}}(L_{\nu}|z)$ that evolves with redshift, the source counts are given by
\begin{align}
    S^2n(S)=\frac{c}{4\pi\ln(10)}&\int_0^{t_{L,\max}}\int_{-\infty}^{\infty}[\,p(\alpha|z)\,\cdot \notag\\
    &u_{\mathrm{dex}}(L_{\nu}|z)(1+z)^{\alpha}\,]d\alpha dt_L,
\end{align}
where $t_L$ is the lookback time and $t_{L,\max}\approx 13.46\,\mathrm{Gyr}$. Both AGN and SFGs are assumed to have different spectral index distributions. We keep the same AGN spectral index distribution as \citetalias{Matthews_2021b}, which includes a double Gaussian with a dominant steep spectrum and less-dominant flat spectrum source population components. The SFG spectral index distribution is determined by the empirical observations of \cite{1990ApJ...357...97C}, where the ratio of the nonthermal to thermal flux density for the typical local star forming galaxy is given by:
\begin{equation}
    \frac{S_n}{S_t} \approx 10\left(\frac{\nu}{\mathrm{GHz}}\right)^{-0.7}.
\end{equation}

\citetalias{Matthews_2021b} model the spectral index distribution of SFGs as a single Gaussian where the mean spectral index at redshift $z$ is the flux-weighted average of the thermal $\alpha_t=-0.1$ and nonthermal $\alpha_n=-0.8$ spectral indices at the corresponding rest-frame frequencies. However, this tends to overestimate the flattening in the spectral index between $1.4\,\mathrm{GHz}$ and $1.4(1+z)\,\mathrm{GHz}$ at higher redshifts due to spectral curvature between the observed and rest frame frequencies and the spectral index being flatter at $1.4(1+z)\,\mathrm{GHz}$ by nature of the average radio spectral energy distribution (SED) of SFGs. We instead compute the effective spectral index between the observed frequency $\nu=1.4\,\mathrm{GHz}$ and the rest-frame frequency $\nu_0=\nu(1+z)$ as
\begin{equation}
    \alpha_{\mathrm{eff}} = \frac{1}{\log(1+z)}\log\left(\frac{\nu_0^{-0.1}+10\nu_0^{\alpha_n}}{\nu^{-0.1}+10 \nu^{\alpha_n}}\right).
\end{equation}
Assuming that $\langle \alpha_n\rangle = -0.8$ with $\sigma_{\alpha_n}=0.17$ as done in \citetalias{Matthews_2021b}, we approximate the spectral index distribution as a Gaussian with mean $\alpha_{\mathrm{eff}}$ and standard deviation $\sigma_{\alpha_{\mathrm{eff}}}=\left|\partial\alpha_{\mathrm{eff}}/\partial\alpha_n\right|\sigma_{\alpha_n}$ to first order.

\subsection{Conversion to SFRD}
The corresponding luminosity density functions determined by the luminosity and density evolutions $f(z)$ and $g(z)$, respectively, can be integrated at each redshift to obtain an SFRD. This is done by using the parameter $q_{\mathrm{FIR}}$ defined in Equation~1 to convert radio luminosity densities to FIR luminosities and then a standard IMF (assumed here to be Salpeter) to convert the FIR luminosities to SFRs. \citetalias{Matthews_2021b}, using the SFG sample of \cite{2019ApJ...872..148C}, find the local radio-FIR correlation to be luminosity dependent
\begin{equation}
    q_{\mathrm{FIR}} = \begin{cases}
        2.69-0.147[\log(L_{1.4}) - 19.1] &\log(L_{1.4}) < 22.5 \\
        2.19 & \log(L_{1.4}) \geq 22.5,
    \end{cases}
\end{equation}
which is consistent with \cite{2021AA...647A.123D} who find an increasing $q_{\mathrm{FIR}}$ with smaller SFG stellar masses, the origin of which is still not certain and subject to future studies.

Since \citetalias{2014ARA&A..52..415M} convert integrated TIR luminosities from $8$--$1000\,\mathrm{\mu m}$ to SFRs using a constant factor $\kappa_{\mathrm{TIR}} = 4.5\times10^{-44}\,\mathrm{M_{\odot}\,yr^{-1}\,erg^{-1}\, s}$ for a Salpeter IMF, but we use an FIR luminosity--based conversion, we adopt the empirical correction factor of $\langle L_{\mathrm{TIR}}/L_{\mathrm{FIR}}\rangle\approx 2$ to $\kappa_{\mathrm{TIR}}$ to convert FIR luminosities to SFRs, consistent with \cite{2003ApJ...586..794B} who find an average ratio ${\sim}1.91$ and assumed by \cite{2009ApJ...706..482M} to match the average $q_{\mathrm{FIR}}$ under a 10\% thermal fraction to the average $q_{\mathrm{TIR}}$. We note that this conversion factor may also be slightly larger, with \cite{10.1093/mnras/staf1536} finding $\langle L_{\mathrm{TIR}}/L_{\mathrm{FIR}}\rangle\approx 2.29$; however, adopting this larger factor only increases the radio source count discrepancy by ${\sim}15\%$. These differences can arise from the different ways in which TIR luminosities are derived, either by fitting FIR SEDs or using the IRAS prescription that combines 12, 25, 60, and 100 $\mu \mathrm{m}$ flux densities \citep[e.g., as used by][]{1996ARA&A..34..749S}. The SFRD $\psi(t)$ in $\mathrm{M_{\odot}\, Mpc^{-3}\, yr^{-1}}$ is then given by:
\begin{align}
    \psi(t)=9\times10^{-37}\cdot 3.75\times10^{12}\cdot \notag & \\
    \int_{-\infty}^{\infty}u_{\mathrm{dex}}(L_{\nu}|t)10^{q_{\mathrm{FIR}}(\log L_{\nu})}d\log L_{\nu},
\end{align}
where $u_{\mathrm{dex}}(L_{\nu}|t)$ is in $\mathrm{W\,Hz^{-1}\, Mpc^{-3}\, dex^{-1}}$. 

\section{Source Count Discrepancy under a Nonevolving radio-FIR Correlation}
\citetalias{Matthews_2021b} show that for redshifts $1\lesssim z\lesssim 2$, the radio source counts imply a luminosity and density evolution that is potentially discrepant with UV/IR measurements of the SFRD by a factor of a few. Furthermore, for redshifts above the turnover in the SFRD $z\gtrsim 2$, the radio--derived cosmic SFRD appears to underestimate the UV measurements. To show this discrepancy, in this section, we consider finding the luminosity and density evolution functions that best match the observed cosmic SFRD based on IR and extinction--corrected UV measurements (using Equation~9) and compare the implied source counts from that evolution to the observed radio source counts (using Equation~5).

\subsection{Literature SFRD Measurements}
In this work, we focus on the discrepancy for redshifts $z\lesssim 3$, over which UV and IR-based SFRD measurements tend to agree. Beyond $z{\sim}4$, differences in sample selections, dust corrections, and tracers can lead to substantial discrepancies in the SFRD exceeding $1\,\mathrm{dex}$ even among IR-selected surveys \citep[e,g., see compilations within][]{10.1093/mnras/staf1961, 2026A&A...706A.345K}. This also is consistent to the highest redshifts out to which direct constructions of the radio LF at 1.4 GHz have provided constraints on the SFRD \citep[e.g.,][]{novak_2017, 2024ApJ...966..194M, 2026MNRAS.548ag616T}. In addition to the UV/IR measurements obtained by \citetalias{2014ARA&A..52..415M}, we incorporate other IR-based SFRD measurements from more recent surveys \citep{2019A&A...624A..98W, 2020A&A...643A...8G, 2024A&A...690A..84T, 2024ApJS..275...36F, 2026MNRAS.545f1961L, 2026A&A...706A.345K}, which are based on IR LFs constructed from the Atacama Large Millimeter/submillimeter Array (ALMA) and Herschel data and are broadly consistent with the compilation of \citetalias{2014ARA&A..52..415M}. We note that these are not radio-selected surveys but instead IR-selected surveys, which we use to constrain the evolution in the radio LF. The benefit of combining multiple independent surveys is to achieve a greater statistical constraint on the $g(z)f(z)$ evolution that determines the shape of the SFRD. Measurements are normalized to the Salpeter IMF and the $\kappa_{\mathrm{TIR}}$ luminosity conversion factor used by \citetalias{2014ARA&A..52..415M}.

\begin{figure}
    \centering
    \includegraphics[width=\linewidth]{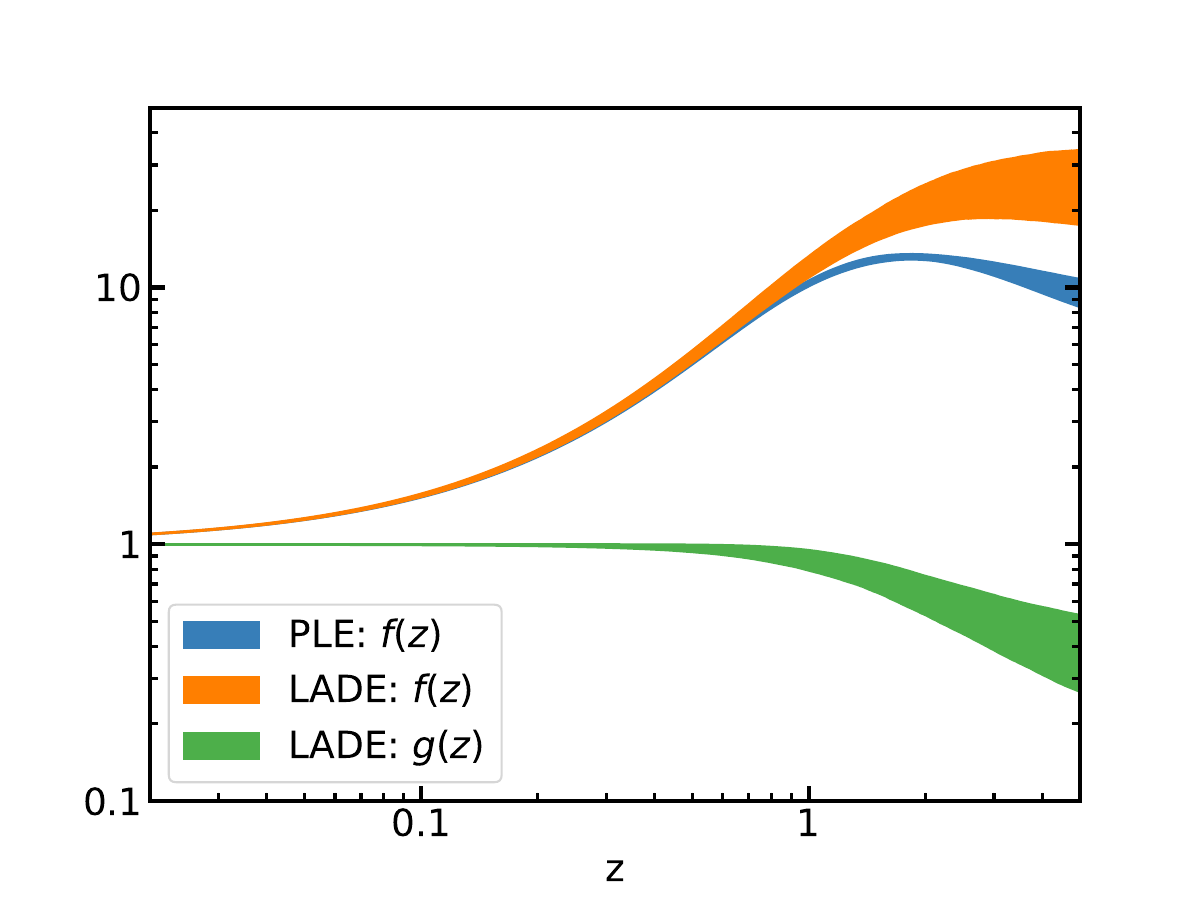}
    \caption{1$\sigma$ intervals in the luminosity $f(z)$ and density $g(z)$ evolution to the local radio LF when fitted to UV/IR SFRD data out to $z=3$ for both the PLE and LADE models. We have extended the figure out to $z=5$ as an extrapolation.}
    \label{fig:fig_1.5}
\end{figure}

\subsection{Luminosity and Density Evolution Priors}
We assume a nonevolving $q_{\mathrm{FIR}}$ with redshift, but maintain the luminosity-dependent parameterization as shown in Equation~8. We use the parameterization of $f(t)$ and $g(t)$ given by M21b,
\begin{equation}
    f(t)=0.5\left[\mathrm{erf}\left(\frac{t-t_f}{\tau_f}\right)+1\right]\exp\left(\frac{t_0-t}{\tau_1}\right),
\end{equation}
resulting in a total of six parameters: $t_{f,g}$ representing the characteristic turn-on time, $\tau_{f,g}$ representing the timescale of the turn-on, and $\tau_{1,2}$ representing the decay timescale of the luminosity $f(t)$ and density $g(t)$ evolutions respectively. Here, $t_0=13.46\,\mathrm{Gyr}$. Because $L_{\star}$ galaxies drop out of samples by $z{\sim}1$, the luminosity and density evolution are degenerate \citep[e.g.,][]{novak_2017, 2022ApJ...941...10V, 2026MNRAS.548ag616T}. We therefore consider the following models, which have been successful at reproducing radio LFs out to $z{\sim}3$:
\begin{itemize}
    \item a pure luminosity evolution (PLE) model, where $g(t)=1$ for all $t$,
    \item and a luminosity and density evolution (LADE) model. Here, we keep the late time density evolution constant $\tau_g=-\infty$ consistent with \citetalias{Matthews_2021b}.
\end{itemize}
The PLE model consists of just three free parameters, while the LADE results in five free parameters.

\subsection{Results}
We use the MCMC package \textsc{emcee} \citep{2013PASP..125..306F} to obtain uncertainties over the luminosity and density evolution parameters and hence the resulting radio source counts for both models. We apply broad uniform priors with each parameter sampled uniformly from $0.1$ to $10\,\mathrm{Gyr}$ and adopt a Gaussian log-likelihood with symmetrized uncertainties. The resulting luminosity and density evolutions are shown in Figure~\ref{fig:fig_1.5} for both models. Direct constructions of radio LFs of SFGs by different surveys have not yet achieved a consensus on the luminosity and density evolution. For example, in the VLA-COSMOS field, \cite{novak_2017} show that a PLE is sufficient to explain observed radio LFs out to $z{\sim}5$, with the luminosity evolution peaking between $2\leq z\leq 3$ at ${\sim}20\times$ higher than $z=0$. Alternatively, \cite{10.1093/mnras/stad1602, 2024A&A...683A.174W} infer a luminosity evolution that may increase out to at least $z\approx 3$ at $\gtrsim100\times$ higher than $z=0$, while densities start to decrease rapidly at around $z{\sim}1$ \citep[see Figure 8 in][]{2024A&A...683A.174W}. The LADE model, however, is generally consistent with observed IR LF evolutions \citep[e.g.,][]{2019A&A...624A..98W, 2020A&A...643A...8G}, with the density evolution remaining weak out to $z\lesssim 1$ and the luminosity evolution increasing by ${\sim}30\times$ out to $z\approx3$. In any case, in Appendix A, we show that the luminosity/density evolution constrained out to $z=3$ is the most relevant for source count agreement near the SFG peak.

\begin{figure}[t!]
    \centering
    \includegraphics[width=\linewidth]{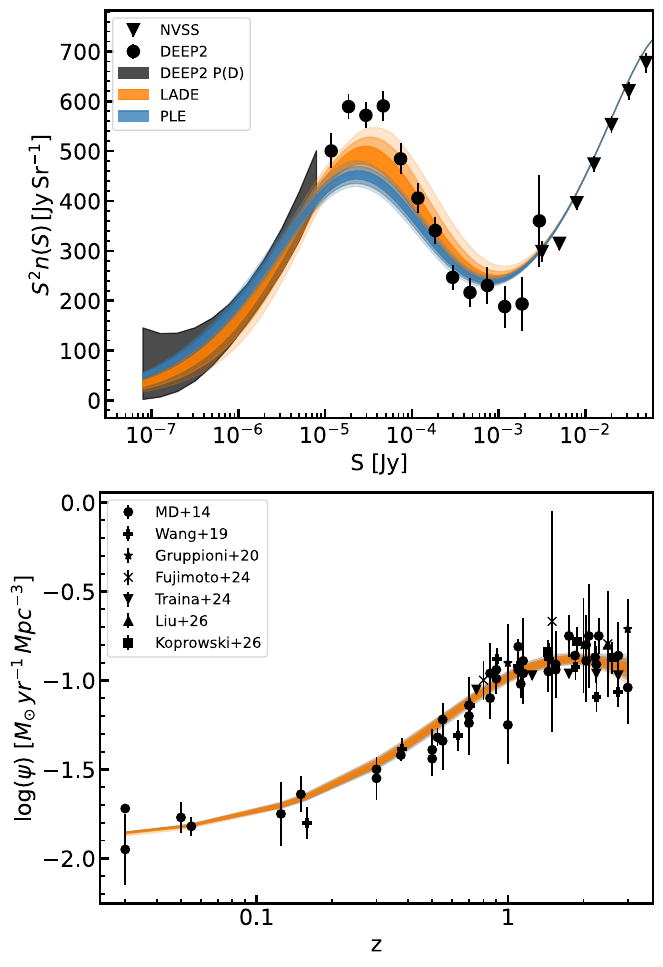}
    \caption{Both a LADE (orange) and PLE (blue) model consistent with the UV/IR SFRD out to $z=3$ (bottom) yield $>3\sigma$ discrepancy in at least one flux density bin near the SFG source count peak ${\sim}30\,\mathrm{\mu Jy}$ (top). The median LF evolution tends to underestimate the radio source counts by ${\sim}20$--$30\%$. In the top panel, we have plotted the modeled counts against observed source counts and their $1\sigma$ uncertainties constrained from the DEEP2 P(D) (black, shaded) and direct counts (circles) and from NVSS direct counts (triangles). In the bottom panel, in addition to the compilation of UV/IR measurements from \citetalias{2014ARA&A..52..415M}, we have also incorporated recent IR measurements from \cite{2019A&A...624A..98W, 2020A&A...643A...8G, 2024A&A...690A..84T, 2024ApJS..275...36F, 2026MNRAS.545f1961L, 2026A&A...706A.345K}. In both panels, we plot the $1,2$, and $3\sigma$ intervals in decreasing opacity.}
    \label{fig:fig2}
\end{figure}

In Figure~\ref{fig:fig2}, we show the resulting fit to the radio source counts and cosmic SFR history. We observe that both the PLE and LADE models tend to underpredict the observed radio source counts near the SFG peak at $S_{1.4}{\sim}30\,\mathrm{\mu Jy}$ by $\gtrsim 3\sigma$. As shown in Appendix A, the greatest contribution to the radio source counts over this range of flux densities comes from radio LFs at $z\approx 0.5$--$2$, where $L_{\star}\approx f(z=1.5)L_{\star,0}\approx 10^{22}\,\mathrm{W\, Hz^{-1}}$, which is the range of redshifts where \citetalias{Matthews_2021b} overestimate the UV/FIR SFRDs. We note, however, that this discrepancy can be slightly reduced assuming $\langle L_{\mathrm{TIR}}/L_{\mathrm{FIR}}\rangle\approx 1.91$ from \cite{2003ApJ...586..794B} or increased assuming a ratio ${\sim}2.29$ from \cite{10.1093/mnras/staf1536}, which holds even in the nonlinear regime of the radio-FIR correlation.

This suggests that the luminosity and/or density evolution under a constant $q_{\mathrm{FIR}}(z)$ is either too weak to reproduce the SFG peak in the radio source counts, or there is sufficient scatter in the conversion from radio to FIR luminosities not accounted for to correct the converted SFRD down from the radio LF. In the following sections, we consider whether combining information from the UV/IR SFRD and the radio source counts can give insight into an evolving $q_{\mathrm{FIR}}(z)$ and its intrinsic scatter.

\section{Redshift-Evolution of the radio-FIR Correlation}
\subsection{Observations of the Radio-FIR Correlation Evolution}
There is some observational evidence that $q_{\mathrm{FIR}}(z)$ decreases mildly out to intermediate redshifts at fixed stellar mass. \cite{2021AA...647A.123D} show that previously derived $q_{\mathrm{FIR}}(z)$ evolutions \citep[e.g.,][]{2017A&A...602A...4D, 2017MNRAS.469.3468C} out to $z{\sim}3$--$4$ are most likely due to selection effects, where the most massive SFGs are preferentially selected out to higher redshift, which are observed to have a smaller $q_{\mathrm{FIR}}$ than the lower mass counterparts. When accounting for stellar mass, \cite{2021AA...647A.123D} show that any redshift evolution is much weaker, with the most massive SFGs exhibiting $q_{\mathrm{FIR}}(z){\sim}(1+z)^{-0.055}$. When fitted across all stellar mass bins down to $M_{\star}{\sim}10^{8}\,\mathrm{M_{\odot}}$, the evolution is even weaker with $q_{\mathrm{FIR}}(z){\sim}(1+z)^{-0.023}.$ IR-selected studies of the radio-FIR correlation, particularly of sub-millimeter galaxies (SMGs), further show evidence of minimal evolution with redshift. SMGs, however, appear to have a significantly lower $q_{\mathrm{FIR}}$, requiring that they have stronger magnetic field strengths compared to normal SFGs at the same redshifts or less dense ISMs to avoid significant cosmic-ray losses \citep{Algera_2020}.

The marginal decline in $q_{\mathrm{FIR}}$ at fixed stellar mass does not necessarily mean a decline in $q_{\mathrm{FIR}}$ at fixed radio luminosity. Given the rapid redshift evolution in the star forming main sequence \citep[e.g.,][]{2014ApJS..214...15S, 2015AA...575A..74S, 2024AA...691A.164K}, SFGs of the same stellar mass will correspond to larger SFRs while SFGs producing the same total radio luminosity may correspond to lower stellar masses. This is observed, for example, in Figure 5 of \cite{2023A&A...679A..47S}, where SFGs of fixed stellar mass correspond to higher radio luminosity SFGs with increasing redshift out to $z=4$. This naturally causes $q_{\mathrm{FIR}}$ to decrease initially at fixed stellar mass. Hence, the rate of decrease in $q_{\mathrm{FIR}}(z)$ at fixed radio luminosity must necessarily be weaker than that at fixed stellar mass, if any. We nevertheless consider a decreasing $q_{\mathrm{FIR}}(z)$ consistent with \cite{2021AA...647A.123D} at fixed radio luminosity as an illustrative example.

\subsection{A Model of the radio-FIR Correlation Evolution} 

Due to increasing cosmic-ray losses at higher redshifts, including inverse Compton scattering from the cosmic microwave background (CMB) and ionization/bremsstrahlung losses in higher density environments, the fraction of thermal emission is expected to rise due to the suppression in the synchrotron emission, particularly for lower mass and luminosity SFGs \citep{2009ApJ...706..482M, 2023A&A...679A..47S, 2024ApJ...975...15Y}. By calibrating $L_{\nu}^{T}\propto L_{\mathrm{FIR}}$, we can express $q= q_0+\log(f_t),$ where $f_t$ is the thermal fraction at 1.4 GHz and $q_0=3.32$ is interpreted to be the calibration near $L_{\star}(z=0)$ SFGs. Therefore, an increasing thermal radio fraction will result in an increasing $q_{\mathrm{FIR}}$ at fixed $q_0$ and hence an increasing SFR per unit radio luminosity generated, thus resulting in the ``breakdown'' of the radio-FIR as correlation as predicted by \cite{2009ApJ...706..482M, 2013AA...556A.142S, 2023A&A...679A..47S, 2024ApJ...975...15Y} at high redshifts. Indeed, stacking of deep MIGHTEE observations have suggested that among the more massive SFGs, $q_{\mathrm{FIR}}$ starts to increase for $z\gtrsim 3$ following an initial decrease, likely due to such redshift-dependent cosmic-ray losses beginning to take over \citep{2025MNRAS.543..507W}. 

Significant work has already been done to model $q_{\mathrm{FIR}}(z)$ evolution either for galaxies of a fixed magnetic field strength \citep[e.g.,][]{2009ApJ...706..482M}, fixed gas-mass surface density \citep[e.g.,][]{2024ApJ...975...15Y}, or stellar mass \citep[e.g.,][]{2023A&A...679A..47S}. However, these models do not indicate how galaxies of the same radio luminosity, a directly observable parameter, at different redshifts are expected to evolve and are thus not compatible with backwards evolutionary models to match radio source counts. Since $L_{\star}$ galaxies at any redshift dominate the radio luminosity density and hence SFRD, it is important to constrain how $q_{\mathrm{FIR}}$ changes for them.

Some existing models of $q_{\mathrm{FIR}}(z)$ derived from synchrotron suppression factors \citep[e.g.,][]{2009ApJ...706..482M, 2024ApJ...975...15Y} ignore the possibility of a changing FIR-to-thermal radio calibration $q_0$ with redshift, which, as discussed in Section 7.3, may dampen the ``breakdown'' in the radio-FIR correlation. By modeling radio emission from first-principles, \cite{2023A&A...679A..47S}, however, are able to take this effect into account, with the caveats that predicted thermal fractions at $L_{\star}(z=0)$ are significantly lower than observed and radio luminous SFGs not initially on the star forming main sequence are not accounted for. Our proposed model, which follows the evolution of $q_{\mathrm{FIR}}(z)$ that is primarily driven by a changing synchrotron suppression factor at fixed radio luminosity, is therefore only approximate.

\subsubsection{Model Description}
We first define the synchrotron suppression factor $f_s=f_s(L_{1.4}, z, \Delta q_{\mathrm{FIR}}, \nu)$, where $L_{1.4}$ is the 1.4\,GHz luminosity at a given redshift $z$, $\Delta q_{\mathrm{FIR}} = q_{\mathrm{FIR}}(L_{1.4},z)-q_{\mathrm{FIR}}(L_{1.4}, z=0)$ is the change in $q_{\mathrm{FIR}}$ relative to a galaxy of the same luminosity at $z=0$, and $\nu$ is the frequency at which the suppression factor is calculated. Assuming that the thermal and nonthermal luminosities are directly correlated to the same SFR quantity by some factors $a_{\nu}$ and $b_{\nu}$ before any suppression $f_s$ \citep[see Equations 10-14 in][]{2011ApJ...737...67M}, we can define
\begin{equation}
\begin{split}
    L_{\nu}(z) &= L_{\nu}^{\mathrm{T}}(z) + L_{\nu}^{\mathrm{NT}}(z)  \\
    &= a_{\nu}\mathrm{SFR}(z) + b_{\nu}\mathrm{SFR}(z)f_s.
\end{split}
\end{equation}
The nonthermal component can be interpreted as that from supernovae accelerating cosmic-rays before encountering various loss mechanisms that would cause its reduction by the time the emission leaves the galaxy. Consequently, the SFR must change to maintain the same radio luminosity. Then, the thermal fraction of a galaxy with spectral luminosity $L_{\nu}$ at redshift $z$ is given by:
\begin{equation}
    f_t=\frac{L_{\nu}^\mathrm{T}(z)}{L_{\nu}^{T}(z)+L_{\nu}^{\mathrm{NT}}(z)} = \frac{a_{\nu}}{a_{\nu}+b_{\nu}f_s}.
\end{equation}
The conditions at $z=0$ set the relationship between $a_{\nu}$ and $b_{\nu}$, where there is no redshift-related synchrotron suppression factor $f_s=1$. There is evidence that lower luminosity SFGs have larger thermal fractions, as either indicated by comparisons to H$\alpha$ luminosities or direct constructions of radio SEDs out to frequencies ${\sim}10$-$30\,\mathrm{GHz}$ \citep{2003ApJ...586..794B, 2012MNRAS.423L.127R, 2017ApJ...836..185T, 2018A&A...611A..55K}. The increase in the radio thermal fraction or, equivalently, synchrotron suppression down to low luminosities could be the driver of the stellar mass (or radio luminosity) dependence on the radio-FIR correlation because $q\propto q_0+\log(f_t)$. A detailed study of this dependence on a large sample of galaxies still remains to be done due to the difficulty of either obtaining dust-corrected total H$\alpha$ luminosities or high frequency radio observations where thermal emission dominates. Defining $c(L_{1.4}, \beta) = (L_{1.4}/L_{\star})^\beta$, where $\log(L_{\star})=21.16$ (in $\mathrm{W\, Hz^{-1}}$), we find that a reasonable approximation to the thermal fraction dependence on total radio luminosity at $z=0$ based on the galaxy sample of \cite{2017ApJ...836..185T} can be obtained by requiring that 
\begin{equation}
b_{\nu}=10c(L_{1.4},\beta)\nu_{\mathrm{GHz}}^{-0.7} a_{\nu},
\end{equation}
with $\beta \approx 0.39$ (see Appendix B), assuming that $L_{\star}$ galaxies have on average ${\sim}10\%$ thermal fractions at 1.4 GHz. From Equation~12, we obtain a thermal fraction depending only on the nonthermal to thermal ratio at $z=0$ and the additional synchrotron suppression factor $f_s$:
\begin{equation}
    f_t= \frac{1}{1+10c(L_{1.4},\beta)\nu_{\mathrm{GHz}}^{-0.7}f_s}.
\end{equation}
Because $f_s$ changes for galaxies of the same luminosity, the corresponding SFR must change as well. Suppose $L_{\nu}(z)=L_{\nu}(0).$ Then relating Equation~11 to that evaluated at $z=0$,
\begin{align}
   \frac{\mathrm{SFR}(z)}{\mathrm{SFR}_0} = \frac{a_{\nu}+b_{\nu}}{a_{\nu}+b_{\nu}f_s} = \frac{f_t(z)}{f_t(0)} \equiv 10^{\Delta q_{\mathrm{FIR}}},
\end{align}
where we have assumed that $\Delta q_{\mathrm{FIR}}$ is driven primarily by changes in thermal radio fractions, and that $\Delta q_0(z)=0$ at fixed radio luminosity \citep[e.g., consistent with the assumptions of][who assumed a fixed $q_0\approx 3.35$ with redshift independent of gas-mass surface density]{2024ApJ...975...15Y}. In Section 7.3, we discuss the impact of this assumption on our inferred $q_{\mathrm{FIR}}(z)$ evolution.

Since $L_{\nu}^{\mathrm{NT}}\propto E^{1-p}(P_{\mathrm{synch}}/P_{\mathrm{loss}})\frac{dE}{d\nu}$ per unit SFR \citep{2009ApJ...706..482M, 2011hea..book.....L}, where $p\approx 2.6$ is the assumed cosmic-ray injection index producing the typical nonthermal spectral index $\alpha_{\mathrm{n}}=-0.8$, $P_{\mathrm{synch}}$ is the synchrotron power, and $P_{\mathrm{loss}}$ is the sum of the energy loss rates of cosmic-rays not contributing to the 1.4\,GHz emission, we take the synchrotron suppression factor (at fixed rest-frame frequency of emission) to be
\begin{equation}
    f_s = \frac{B(z)^{(p-2)/2}P_{\mathrm{synch}}(z)P_{\mathrm{loss}}(0)}{B(0)^{(p-2)/2}P_{\mathrm{synch}}(0)P_{\mathrm{loss}}(z)},
\end{equation}
where the extra factor of the magnetic field strength ratio comes from 1.4\,GHz emission sampling a different part of the cosmic-ray injection spectrum and thus a different cosmic-ray density contributing to the 1.4\,GHz emission. $P_{\mathrm{loss}}$ includes synchrotron, ionization, bremsstrahlung, inverse Compton off the CMB and interstellar radiation fields, and escape \citep[see][for loss timescale equations]{2009ApJ...706..482M}. Calculating these loss timescales requires specifying an ISM density, magnetic field strength, and size prescription across SFGs of different luminosities and at different redshifts. We describe these prescriptions in detail in the following subsections, which are summarized in Table \ref{tab:model_scalings}.

\begin{deluxetable}{llll}
\tablecaption{Model Parameters with Normalization and Evolution\label{tab:model_scalings}}
\tablehead{
\colhead{Parameter} &
\colhead{Normalization (MW)} &
\colhead{Evolution} &
\colhead{Reference}
}
\startdata
ISM Density: $n_{\mathrm{ISM}}$ &
$n_0 = 0.8\,\mathrm{cm^{-3}}$ &
$n_{\mathrm{ISM}} \sim n_0\,\mathrm{SFR}^{1/1.4} R_e^{-17/7}$ &
K98
\\
Magnetic Field Strength: $B$ &
$B_0 = 10\,\mu\mathrm{G}$ &
$B \sim B_0 \left(n_{\mathrm{ISM}}/n_0\right)^{1/6}
\Sigma_{\mathrm{SFR}}^{1/3}$ &
SB13
\\
Radio Continuum Size: $R_e$ &
\nodata &
$R_e \sim (1+z)^{-0.26}$ &
JA21
\\
Escape Scale Height: $h$ &
$h_0 = 3\,\mathrm{kpc}$ &
$h \sim h_0(1+z)^{-0.26}$ &
K18
\\
ISRF Energy Density: $U_{\star}$ &
$U_{\star,0} = U_{B_0}/3$ &
$U_{\star}(z) = U_B(z)/3$ &
YH16
\enddata
\tablecomments{
The normalizations for the ISM density, magnetic field strength, and escape scale height are chosen for MW-like galaxies. The radio continuum size evolution has no initial normalization because it only appears in the ISM density and magnetic field evolutions, which are already normalized to MW values.
}
\tablerefs{
K98: \cite{1998ApJ...498..541K};
SB13: \cite{2013AA...556A.142S};
JA21: \cite{Jiménez-Andrade_2021};
K18: \cite{2018AA...611A..72K};
YH16: \cite{2016MNRAS.457L..29Y}
}
\end{deluxetable}

\subsubsection{Ionization/Bremsstrahlung Losses: ISM Density}

The ionization and bremsstrahlung timescales are dependent on the average ISM density, which we scale to the Milky Way (MW). We estimate this by computing the volume filling factor-weighted density of the various phases of the ISM \citep[molecular gas, cold/warm neutral medium, HII regions, and hot-ionized medium, see Table 1.3 of][]{2011piim.book.....D}. This produces a relevant MW $z=0$ ISM density of $n_{\mathrm{ISM}}=0.8\,\mathrm{cm^{-3}}$ that we scale SFGs of other luminosities to. We note that all phases of the ISM contribute to bremsstrahlung losses, while the neutral and molecular phases will be relevant for ionization losses. As described by \cite{2011hea..book.....L}, bremsstrahlung losses for neutral and ionized hydrogen are consistent to within a factor of ${\sim}2$.

We take $n_{\mathrm{ISM}}\propto M_{\mathrm{gas}}/V\ \propto \Sigma_{\mathrm{gas}}R_e^2/V \propto \Sigma_{\mathrm{gas}}/R_e.$ From the global Kennicutt-Schmidt relation, we take $\Sigma_{\mathrm{gas}}\propto \Sigma_{\mathrm{SFR}}^{1/1.4}$ \citep{1998ApJ...498..541K, 2021ApJ...908...61K}, generally consistent for both normal and starburst SFGs. Here, $R_e$ is the effective size of the radio-emitting region of the SFG. Therefore, we take $n_{\mathrm{ISM}}\propto \mathrm{SFR}^{1/1.4}R_e^{-17/7}$ with $\mathrm{SFR}\propto L_{1.4}10^{\Delta q_{\mathrm{FIR}}}.$ \cite{Jiménez-Andrade_2021, 2021AA...647A.123D} show that radio continuum sizes are observed to decrease much more weakly with redshift relative to optical/UV sizes. We take the radio continuum size evolution of \cite{Jiménez-Andrade_2021} for normal SFGs evolving as $R_e{\sim}(1+z)^{-0.26}.$ We assume negligible radio continuum size dependencies across SFGs of different luminosities at a fixed redshift, as further suggested by \cite{Jiménez-Andrade_2021} and assumed in previous works \citep[e.g.,][]{2023A&A...679A..47S}. Normalizing to a MW radio luminosity of $L_{1.4}^{\mathrm{MW}}\approx 2\times10^{21}\,\mathrm{W\, Hz^{-1}}, $ we compute ISM densities as:
\begin{equation}
\begin{split}
    n_{\mathrm{ISM}}&\approx 0.8\left(\frac{L_{1.4}}{L_{1.4}^{\mathrm{MW}}}\right)^{1/1.4}\left(10^{\Delta q_{\mathrm{FIR}}(z, L_{1.4})}\right)^{1/1.4} \\
    \times &\left(10^{q_{\mathrm{FIR}}(L_{1.4})-q_{\mathrm{FIR}}(L_{1.4}^{\mathrm{MW}})}\right)^{1/1.4}(1+z)^{(0.26 \cdot 17/7)}\,\mathrm{cm^{-3}},
\end{split}
\end{equation}
where we have accounted for an additional correction factor on $q_{\mathrm{FIR}}$ relative to the MW due to the local luminosity dependence on $q$, thus self-consistently using the FIR emission as a tracer for the relevant SFR quantity determining the ISM density.

\subsubsection{Synchrotron Losses: Magnetic Field Strength}
Following \cite{2013AA...556A.142S, 2023A&A...679A..47S}, we assume that the magnetic field strength is proportional to the turbulent kinetic energy density driven by the energy injected into the ISM by supernovae, implying that $B\propto \Sigma_{\mathrm{SFR}}^{1/3}n_{\mathrm{ISM}}(z)^{1/6}.$ We assume that the magnetic fields in galaxies are saturated throughout most of cosmic history; it is expected that magnetic field saturation via seed turbulent dynamos can take $\mathcal{O}(10\,\mathrm{Myr})$ \citep{2013A&A...560A..87S}. We therefore scale the magnetic field strength to the typical MW value of $B_0\approx 10\,\mathrm{\mu G}$ as:
\begin{equation}
\begin{split}
    B &\approx 10\left(\frac{n_{\mathrm{ISM}}(L_{1.4}, z)}{0.8\,\mathrm{cm^{-3}}}\right)^{1/6} \left(10^{\Delta q_{\mathrm{FIR}}(z, L_{1.4})}\right)^{1/3} \\
    &\times\left(\frac{L_{1.4}}{L_{1.4}^{\mathrm{MW}}}\right)^{1/3} \left(10^{q_{\mathrm{FIR}}(L_{1.4})-q_{\mathrm{FIR}}(L_{1.4}^{\mathrm{MW}})}\right)^{1/3} \\
    &\times (1+z)^{(0.26\cdot2/3)}\,\mathrm{\mu G}.
\end{split}
\end{equation}

\subsubsection{Inverse Compton Losses}
We incorporate inverse Compton losses from the CMB with the energy density scaling as $U_B(z) = (4.2\times10^{-13})(1+z)^{4}\,\mathrm{erg/cm^3}$. We also incorporate inverse Compton losses from the interstellar radiation field, taking the energy density to be $U_{\star}\approx 1/3\, U_B$, based on observations of local SFGs and starbursts \citep{2016MNRAS.457L..29Y}.

\subsubsection{Escape Losses: Cosmic-Ray Escape Scale Height}
Determining the cosmic-ray escape loss timescale is the escape scale height. \cite{2018AA...611A..72K} observe typical synchrotron scale heights of $h=1.5\,\mathrm{kpc}$. Assuming equipartition between the cosmic-ray energy density and the magnetic field strength implies a cosmic-ray escape scale height of ${\sim}3\,\mathrm{kpc}$ assuming a nonthermal spectral index of $\alpha_{\mathrm{NT}}=-0.8$ \cite{2009ApJ...706..482M}. We therefore adopt an initial escape scale height of $h_0=3\,\mathrm{kpc}$. As done by \cite{2023A&A...679A..47S}, we assume that the scale height evolves with the effective radius evolution in the same way across all SFGs of different luminosities: $h_{\mathrm{esc}}{\sim}(1+z)^{-0.26}.$

\subsubsection{Model Evaluation}
Because $f_s$ explicitly depends on $\Delta q_{\mathrm{FIR}}$ through the magnetic field strength, $\Delta q_{\mathrm{FIR}}$ can only be solved numerically as a fixed--point problem:
\begin{equation}
    \frac{f_t(z, L_{1.4}, \Delta q_{\mathrm{FIR}}, \nu)}{f_t(0, L_{1.4}, 0 , \nu)} = 10^{\Delta q_{\mathrm{FIR}}}.
\end{equation}
We use \textsc{scipy}'s \textsc{brentq} routine to numerically compute $\Delta q_{\mathrm{FIR}}$ over a grid of spectral luminosities and redshifts, bracketing $\Delta q_{\mathrm{FIR}}\in [-3, 1]$, at three (one) orders of magnitude below (above) the $z=0$ thermal fraction. Over this range, we confirm that there is a unique $\Delta q_{\mathrm{FIR}}$ solution.

\begin{figure}[t!]
    \centering
    \includegraphics[width=\linewidth]{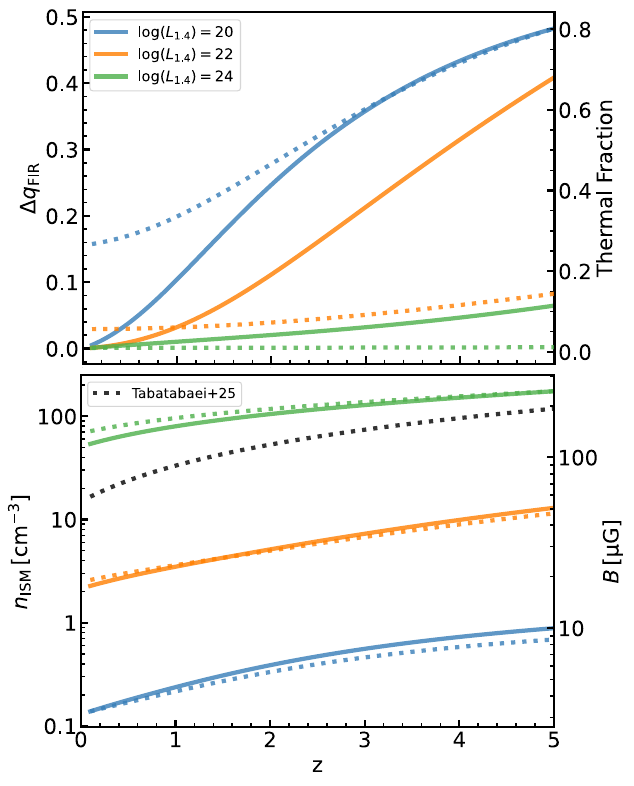}
    \caption{Top: The evolution in $\Delta q_{\mathrm{FIR}}=q_{\mathrm{FIR}}(z)-q_{\mathrm{FIR}, 0}$ (solid) and thermal fraction $f_t$ (dotted) as a function of redshift and different total 1.4 GHz radio luminosities. Bottom: The $\Delta q_{\mathrm{FIR}}$ evolution is driven by an evolution in the average ISM density (solid) and magnetic field strength (dotted). We note that the magnetic field strengths at $10^{24}\,\mathrm{W\, Hz^{-1}}$ tend to overestimate the equipartition magnetic field strengths observed by \citep[][black, dotted]{Tabatabaei_2025} over $1.5\lesssim z\lesssim 3.5$ by a factor of a few, which is largely dependent on the magnetic field normalization set for the Milky Way.}
    \label{fig:fig3}
\end{figure}

In Figure \ref{fig:fig3}, we show how $f_t(z)$, $q_{\mathrm{FIR}}(z)$, $n_{\mathrm{ISM}}(z)$, and $B(z)$ evolve under our prescription. The model magnetic field strength evolution is weaker and normalized higher by a factor of a few compared to the equipartition magnetic field strength evolution observed by \cite{Tabatabaei_2025} over the redshift range $1.5 < z < 3.5$ for SFGs in the luminosity range $23.5\lesssim \log(L_{1.4})\lesssim24.5$, which exhibit a distribution $100 \lesssim (B_{\mathrm{eq}}/\mathrm{\mu G})\leq 200$. We note that the evolution observed by \cite{Tabatabaei_2025} is likely driven by a redshift selection effect as more radio luminous SFGs are represented at higher redshifts. The normalization is also sensitive to the Milky Way magnetic field normalization; for example, reducing the normalization to $B\approx 6\,\mathrm{\mu G}$, consistent with equipartition estimates in the local neighborhood \citep{10.1007/1-4020-2620-X_57} can reduce the normalization mismatch, at the cost of $\Delta q_{\mathrm{FIR}}$ breaking down earlier. Furthermore, as noted by \citep{2005AN....326..414B}, when other cosmic-ray loss processes dominate, the equipartition magnetic field strength may underestimate the true magnetic field strength due to the changing cosmic-ray proton to electron number density ratio.

Over the relevant SFG radio luminosities, we find that the combination of the Kennicutt-Schmidt relation in determining ISM densities and the supernovae turbulence model of \cite{2013AA...556A.142S} produces an approximate scaling $n_{\mathrm{ISM}}(z)\propto B(z)^{2}$ at fixed radio luminosity. This relation is shallower than empirical Davis Chandrasekhar Fermi (DCF)-based estimates of the relation, which find $n_{\mathrm{ISM}}\propto B^{1/0.26}$ \citep[][]{10.1093/mnras/staf901} in the diffuse phases of the ISM with $10^{-3}\lesssim n/\,\mathrm{cm^{-3}}\lesssim10^{5}$.

Maintaining Equation 17 for determining the ISM density, we find that adopting the steeper relation based on DCF observations leads to significantly weaker magnetic field strengths. Consequently, $\Delta q_{\mathrm{FIR}}(z=3)\approx 0.4, 0.45, 0.3\,\mathrm{dex}$ for $L_{1.4}=10^{20}, 10^{22}, 10^{24}\,\mathrm{W\, Hz^{-1}},$ respectively. While $\Delta q(z)$ for SFGs fainter than $L_{\star}(z=0)$ remains generally consistent with our fiducial model, higher luminosity SFGs see a stronger break-down in the radio-FIR correlation relative to that produced by the weaker magnetic field scaling with ISM density. This is because while escape and CMB inverse Compton losses remain dominant relative to synchrotron, ionization, and bremsstrahlung losses in less luminous SFGs, ionization and bremsstrahlung losses start to dominate in more luminous SFGs relative to synchrotron under the weaker magnetic field strengths.

For example, for $L_{1.4}=10^{20}\,\mathrm{W\, Hz^{-1}}$, under the DCF scaling, $P_{\mathrm{synch}}/(P_{\mathrm{CMB}}+P_{\mathrm{esc}})\approx 0.3,0.03$, while under our fiducial model, $P_{\mathrm{synch}}/(P_{\mathrm{CMB}}+P_{\mathrm{esc}})\approx 0.1,0.02$ at $z=0,3$ respectively. Furthermore, in any scaling, $(P_{\mathrm{CMB}}+P_{\mathrm{esc}})\gtrsim 10(P_{\mathrm{ion}}+P_{\mathrm{brem}})$ over this redshift range, so ionization and bremsstrahlung remain insignificant to the cosmic-ray losses. On the other hand, for $L_{1.4}=10^{24}\,\mathrm{W\, Hz^{-1}}$, under the DCF scaling, $P_{\mathrm{synch}}/(P_{\mathrm{ion}}+P_{\mathrm{brem}})\approx 0.15, 0.06$, while under our fiducial scaling, $P_{\mathrm{synch}}/(P_{\mathrm{ion}}+P_{\mathrm{brem}})\approx 1.1, 0.8$ at $z=0, 3$, respectively. We note that in the DCF scaling, the more luminous SFGs start with significantly weaker $B\approx 30\,\mathrm{\mu G}$ growing to only ${\sim}40\,\mathrm{\mu G}$ by $z=3$. Similarly, the combined ionization and bremsstrahlung powers exceed that of CMB inverse Compton and escape in both cases by at least $10\times$ over this redshift range. Therefore, in the DCF scaling, the steeper rise of $\Delta q(z)$ at high radio luminosities can be attributed to weaker magnetic field strengths. Ultimately, differences in the scaling between ISM density and magnetic field strength could arise to differences between resolved and global properties of SFGs.

\begin{figure}[t!]
    \centering
    \includegraphics[width=\linewidth]{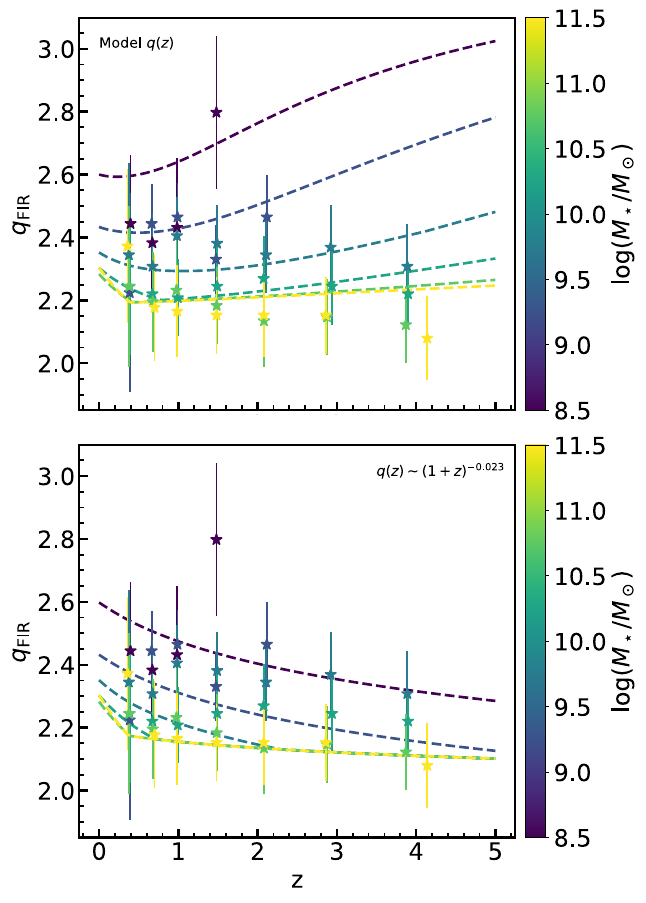}
    \caption{Top: $q_{\mathrm{FIR}}$ evolution at fixed stellar mass as implied by our model and the redshift evolution in the main sequence from \cite{2015AA...575A..74S}. Bottom: the evolution at fixed stellar mass if $q_{\mathrm{FIR}}(z)$ followed the same implied evolution at fixed radio luminosity as observed at fixed stellar mass, which produces a stronger redshift evolution than generally observed, particularly at lower stellar masses. Overlaid in both panels are $q_{\mathrm{FIR}}$ measurements (stars) from \cite{2021AA...647A.123D} shifted down by $0.3\,\mathrm{dex}$ to represent FIR rather than TIR luminosities.}
    \label{fig:fig3.5}
\end{figure}

In Figure \ref{fig:fig3.5}, we show how $q_{\mathrm{FIR}}(z)$ evolves in different stellar mass bins under our prescribed model and separately under the assumption that $q_{\mathrm{FIR}}(z){\sim}(1+z)^{-0.023}.$ This is done by computing an SFR for each stellar mass and redshift bin using a star formation main sequence relation \citep[e.g.,][]{2015AA...575A..74S} and interpolating the corresponding radio luminosity and hence $\Delta q_{\mathrm{FIR}}(z, L_{1.4})$ associated with that SFR. Following \cite{2023A&A...679A..47S}, before converting the main sequence SFR at a given redshift to an associated $\Delta q_{\mathrm{FIR}}$, we adjust it to an IR SFR based on the IRX--$M_{\star}$ relation, which indicates the amount of SFR IR emission does not trace due to the presence of observed UV emission that did not get reprocessed into IR emission. We use the relation by \cite{10.1093/mnras/stu896, 10.1093/mnras/stt1960}, which may not evolve strongly with redshift out to $z{\sim}3$ and potentially beyond \citep[e.g.,][and references therein]{2020ApJ...902..112B}, although some other studies suggest a moderate evolution at the high mass end \citep[e.g.,][]{10.1093/mnras/stad3999}. The correction factor by which IR emission underestimates the SED-derived SFRs in the SFR main sequence is given by:
\begin{equation}
    f_{\mathrm{TIR}} = \frac{\kappa_{\mathrm{TIR}}}{\kappa_{\mathrm{UV}}}\frac{\mathrm{IRX}}{1+\mathrm{IRX}\frac{\kappa_{\mathrm{TIR}}}{\kappa_{\mathrm{UV}}}},
\end{equation}
where $\kappa_{\mathrm{TIR}} = 1.73\times10^{-10}\,\mathrm{M_{\odot}\,yr^{-1}\,L_{\odot}^{-1}}$ and $\kappa_{\mathrm{UV}}=2.2\times10^{-10}\,\mathrm{M_{\odot}\,yr^{-1}\,L_{\odot}^{-1}}$ are the standard conversion factors used by \citetalias{2014ARA&A..52..415M}. Based on \cite{10.1093/mnras/stu896, 10.1093/mnras/stt1960},
\begin{equation}
    \log\mathrm{IRX} = 0.71\left(\log(M/M_{\odot})-10.35\right) + 1.32,
\end{equation}
which is assumed fixed with redshift. Thus, the associated main sequence SFR is first multiplied by $f_{\mathrm{TIR}}$ before finding the associated radio luminosity and $\Delta q_{\mathrm{FIR}}$. Since \cite{2015AA...575A..74S} uses Salpeter IMFs, we adjust the Chabrier masses of \cite{2021AA...647A.123D} to Salpeter masses by adding 0.2 dex. Given that the IRX relation in Equation 21 and the SFR main sequence relation from is based on the Chabrier IMF, we input the mass bins from \cite{2021AA...647A.123D} directly to obtain $f_{\mathrm{TIR}}$.

While the the $q_{\mathrm{FIR}}(z)$ evolution at fixed radio luminosity shows no decrease at low redshifts, the evolution at fixed stellar mass shows evidence of a slight initial decrease across all mass bins, akin to the observations of \cite{2021AA...647A.123D}. Due to the redshift evolution in the star formation main sequence, SFGs of the same mass at higher redshifts have larger SFRs and hence produce more radio emission, thus initially moving down the $q_{\mathrm{FIR}}(z)$ tracks to higher radio luminosities. At higher redshifts, non-synchrotron cosmic-ray losses take over and thus $q_{\mathrm{FIR}}$ still breaks down consistent with \cite{2023A&A...679A..47S}. However, if the same redshift-evolution is applied at fixed radio luminosity, then the resulting redshift evolution at fixed stellar mass is much stronger than observed, particularly at lower stellar masses. Thus, despite the assumption of $\Delta q_0(z)=0$, the $q_{\mathrm{FIR}}(z)$ model driven by thermal fractions provides a better agreement with the stellar mass-dependent radio-FIR correlation.

Naturally, a $q_{\mathrm{FIR}}(z)$ that increases at fixed radio luminosity will produce source counts that may further increase the source count discrepancy because it requires a less strong luminosity evolution to match the SFRD. In this case, an intrinsic scatter correction to the SFRD is necessary. This is further discussed in Sections 5 and 6.

\subsubsection{Spectral Index Evolution}
Because $f_s$ is frequency dependent, the average spectrum in Equation~6 will be modified as a function of redshift and radio luminosity, resulting in the mean spectral index needed for 1.4\,GHz spectral luminosity $k$-corrections being modified. The mean spectral index is taken to be the same as Equation~7, but with the additional suppression factor $f_s(\nu)$ and luminosity-dependent thermal fraction factor $c(L_{1.4}, \beta)$:
\begin{equation}
    \alpha_{\mathrm{eff}} = \frac{1}{\log(1+z)}\log\left(\frac{\nu_0^{-0.1}+10 c(L_{1.4},\beta)f_s(\nu_0)\nu_0^{-0.8}}{\nu^{-0.1}+ 10c(L_{1.4},\beta) f_s(\nu)\nu^{-0.8}}\right).
\end{equation}

\begin{figure}[t!]
    \centering
    \includegraphics[width=\linewidth]{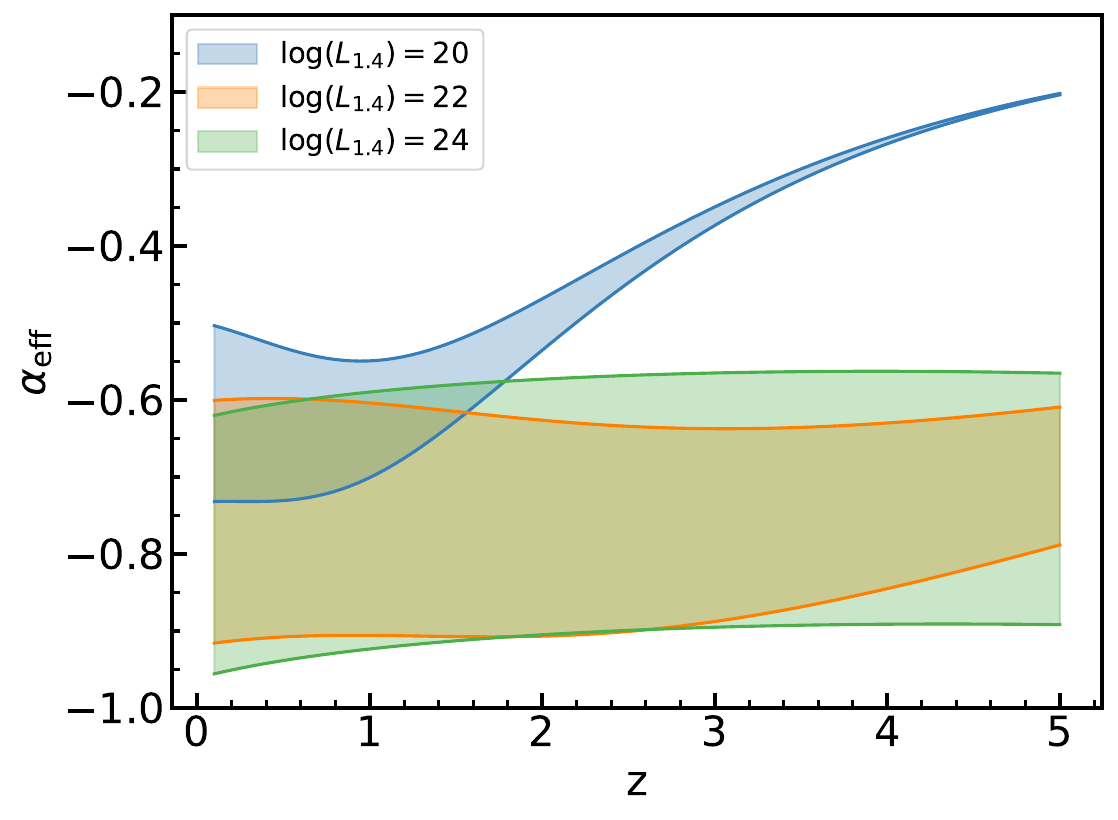}
    \caption{Higher luminosity SFGs retain a relatively consistent spectral index for $k$-correction (due to observing at 1.4 GHz) out to $z=5$ due to thermal fractions at 1.4 GHz evolving weakly. Lower luminosity SFGs rapidly approach the thermal free-free spectral index of $\alpha_t=-0.1$ at higher redshifts.}
    \label{fig:fig4}
\end{figure}

In Figure \ref{fig:fig4}, we plot the redshift evolution in the spectral index distribution for SFGs of different luminosities assumed for $k$-correction. For the more luminous SFGs, the spectral index between $\nu=1.4\,\mathrm{GHz}$ and $\nu_0=1.4(1+z)\,\mathrm{GHz}$ tends to remain consistent across redshift due to the small thermal fractions. The lower luminosity SFGs exhibit a spectral index that rapidly flattens at high redshifts as the thermal component takes over and synchrotron emission gets suppressed.

\section{Effect of Intrinsic Scatter}
In our models that derive the cosmic SFRD from radio source counts, we require a conversion from radio LFs to FIR LFs, which will be affected by the intrinsic scatter in $q_{\mathrm{FIR}}(z)$. If the SFR ultimately determines the radio luminosity, then the resulting radio LF will be broadened at the high-luminosity end leading to a shallower exponential cutoff compared to the intrinsic SFR function \citep[resulting from Eddington Bias,][]{10.1093/mnras/73.5.359}. This has been considered previously by \cite{10.1093/mnras/stad1602} when converting LOFAR deep field 150\,MHz LFs directly to SFR functions. However, at 1.4 GHz, differences in ISM environments across galaxies of the same radio luminosity will lead to scatter in the fraction of nonthermal radio emission to account for the total radio luminosity. Scatter is typically not accounted for when integrating IR and UV LFs to obtain SFRDs (e.g., as in \citetalias{2014ARA&A..52..415M}), so we only account for it going from radio LFs to FIR LFs based on the scatter in $q_{\mathrm{FIR}}(z)$. 

Empirical comparisons between radio and FIR LFs suggest that the FIR LFs may be steeper at the high-luminosity end. In particular, \cite{2013MNRAS.432...23G} fit the same parametric form (Equation~4) to the TIR LF constructed at $0<z<0.3$, finding a $\sigma=0.5$ for spiral galaxies and an even steeper $\sigma=0.35$ for starburst galaxies, compared to the locally derived radio $\sigma\approx 0.66$. A slope $\sigma\approx 0.5$ was subsequently fixed for higher redshift TIR LFs. However, we note that this may be subject to cosmic variance effects; the highest luminosity SFGs have smaller counts and therefore poorer statistics in the LFs, which determine the high-luminosity slope. Defining $x=\log(L_{1.4})$ and $y=\log(\mathrm{SFR})$, the SFR function is such that
\begin{equation}
    \rho_{\mathrm{dex}}(x)=\int\rho_{\mathrm{dex}}(y)P(x|y)dy,
\end{equation}
where $\rho_{\mathrm{dex}}(y)$ is the SFR function and $P(x|y)$ is a Gaussian evaluated at the mean of the transformation between the variables $x$ and $y$
\begin{equation}
    y = x + \log(9\times10^{-37}\times 3.75\times10^{12})+q_{\mathrm{FIR}}(x)
\end{equation}
with standard deviation given by the intrinsic scatter $\sigma_q$ transforming between radio and FIR luminosities.

We approximate the intrinsic FIR LF (and consequently SFR function) to have the same functional form as the radio LF. Because the convolution cannot be inverted analytically, we use an MCMC to sample the parameters of the intrinsic SFR function and perform the Gaussian convolution to match the observed radio LF. We utilize a Gaussian likelihood between the model-convolved radio LF and the observed radio LF combined from multiple surveys, as described in Section 2.

\begin{figure}[t!]
    \centering
    \includegraphics[width=\linewidth]{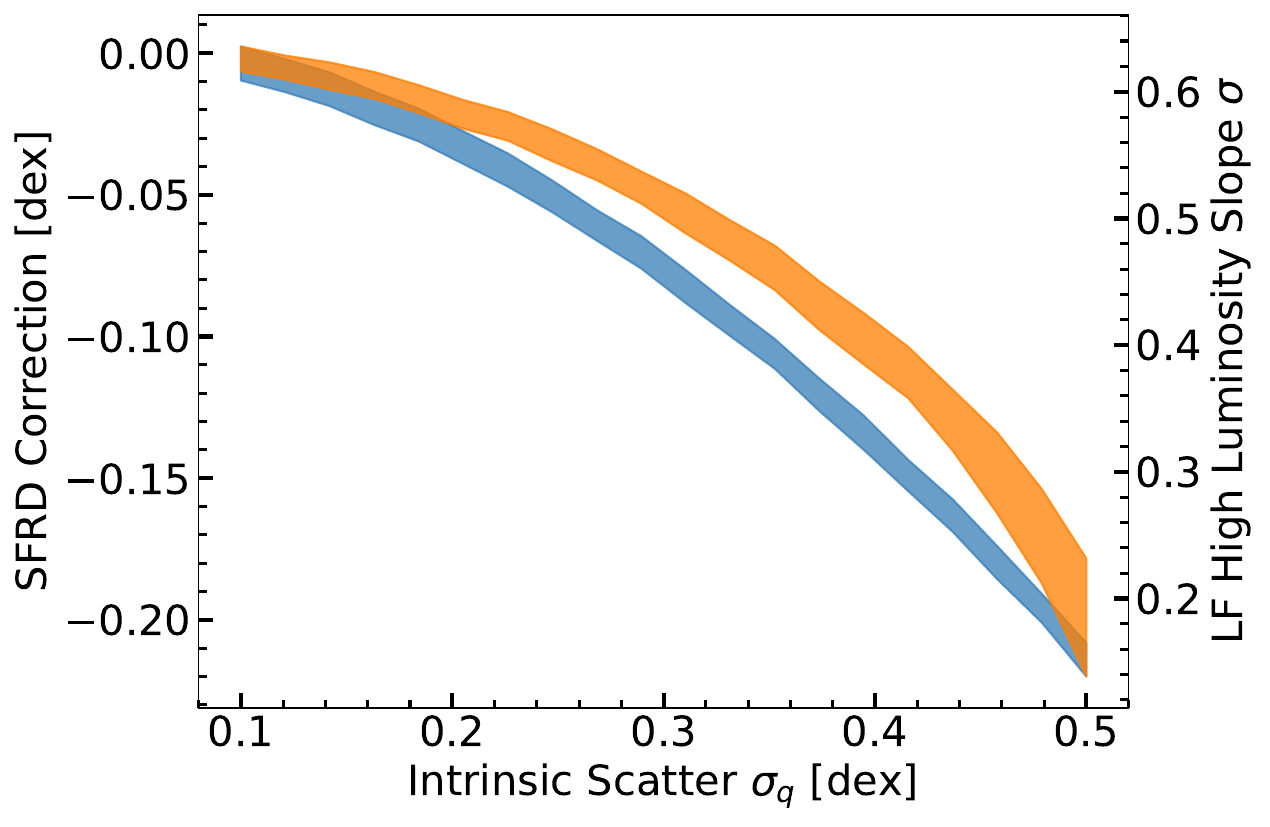}
    \caption{A larger intrinsic scatter $\sigma_q$ produces a steeper intrinsic SFR function slope (orange) and consequently a larger correction to the SFRD (blue). The shaded regions indicate the $1\sigma$ interval derived from the posterior samples.}
    \label{fig:fig5}
\end{figure}

In Figure~\ref{fig:fig5}, we show the level of correction to the SFRD for the $z=0$ radio LF for different choices of scatter $0.1\leq\sigma_q\leq 0.5$ and the corresponding high-luminosity slope $\sigma$. We find that a scatter $\sigma_q\approx 0.3$ is consistent with producing an FIR high-luminosity slope of $\sigma\approx 0.5$, which we take to be the fiducial value. The level of correction imparted on the SFRD is ${\sim}0.08\,\mathrm{dex}$, which provides a statistically significant ${\sim}20$-$30$\% change to the $S^2n(S)$ counts, as further discussed in Sections~6 and 7. 

While this correction is larger than that found by \cite{10.1093/mnras/stad1602}, we note that radio LFs constructed at 150 MHz trace predominantly nonthermal emission and are likely to have a different high-luminosity slope and scatter compared to that of radio LFs constructed at 1.4 GHz. Hence, a direct comparison on the scatter corrections cannot be made. Additionally, our chosen intrinsic scatter is generally larger than that inferred in the local radio luminosity- or stellar mass-dependent radio-FIR correlation \citep[e.g., ${\sim}0.15\,\mathrm{dex}$ and ${\sim}0.2\,\mathrm{dex}$ by][respectively]{2021MNRAS.504..118M, 2021AA...647A.123D} or the linear radio-FIR correlation \citep[e.g., $0.26\,\mathrm{dex}$ by][]{2003ApJ...586..794B}. We note that analyses at higher redshifts assume no intrinsic variation in the spectral index for $k$-corrections. SFGs are observed to show significant variation in the nonthermal spectral index constrained using ${\sim}\mathrm{GHz}$ rest-frame flux densities \citep{Tabatabaei_2025} and in the low frequency spectral index \citep[e.g., from 150 MHz to 1.4 GHz observed,][]{2017MNRAS.469.3468C}, which could reflect different ISM conditions, thermal radio contributions, and even cosmic-ray injection energy distributions. These differences may be more significant at higher redshifts, causing the scatter to be higher. As discussed in Section~7, we find that a larger scatter is required to reproduce the source count peak under a nonevolving radio-FIR correlation or our fiducial model, in contrast to \cite{2021AA...647A.123D}.

Because our model $q_{\mathrm{FIR}}(z)$ is luminosity dependent, the transformation defined in Equation~24 will change with redshift, resulting in a redshift--dependent SFRD correction and slope $\sigma$.

\section{Results}
To evaluate how evolving $q_{\mathrm{FIR}}(z)$ or corrections to the SFRD arising from intrinsic scatter affect the radio source count agreement, we perform the same procedure as Section~3, finding the evolutionary parameters that best match the observed UV/IR SFRD history out to $z=3$ and then comparing to the source counts. We consider five models using both priors on radio LF evolution to test the impact of an evolving $q_{\mathrm{FIR}}(z)$ and intrinsic scatter in the radio-FIR correlation on the agreement with the radio source counts.

\begin{figure}
    \centering
    \includegraphics[width=\linewidth]{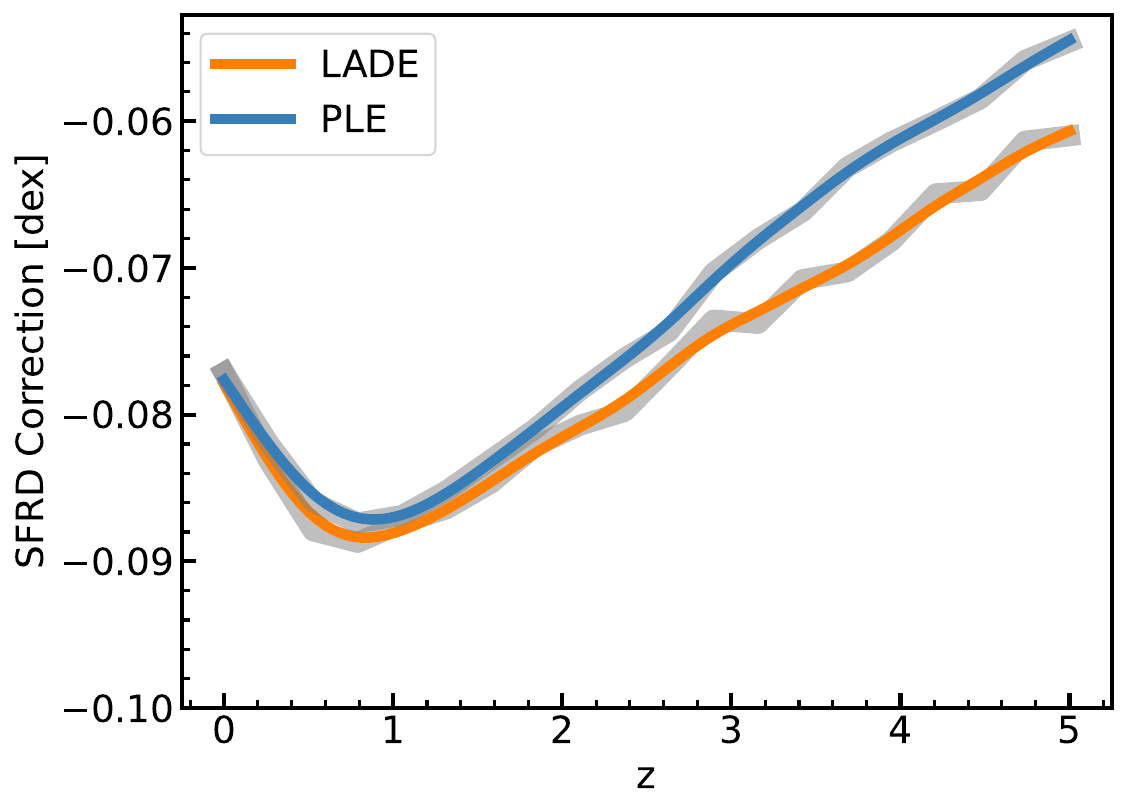}
    \caption{Despite the assumption of a constant scatter $\sigma_q=0.3\,\mathrm{dex}$, the luminosity dependence on $q_{\mathrm{FIR}}(z)$ produces an SFRD correction slightly dependent on redshift, which is interpolated over using a smoothing spline (gray).}
    \label{fig:fig6}
\end{figure}

\begin{enumerate}
    \item Model 1: Motivated by $\sigma_q=0.3\,\mathrm{dex}$ reproducing the high-luminosity slope of the $z=0$ TIR LF of \cite{2013MNRAS.432...23G} and being an upper bound to the observed local scatter in $q_{\mathrm{FIR}}$, we adopt a correction factor of ${\sim}0.08\,\mathrm{dex}$ without any $q_{\mathrm{FIR}}(z)$ evolution to test whether intrinsic scatter alone is sufficient to reduce the source count discrepancy.
    \item Model 2: We modify Equation~8 and incorporate a redshift evolution $q_{\mathrm{FIR}}{\sim}(1+z)^{-0.023}$ at fixed radio luminosity consistent with \citep{2021AA...647A.123D} to independently test whether a monotonically decreasing $q_{\mathrm{FIR}}(z)$ can explain the radio source counts.
    \item Model 3: Same as Model 2, but we incorporate the additional redshift-independent intrinsic scatter correction factor  $0.08\,\mathrm{dex}$ to test the combination of both effects.
    \item Model 4: We apply our fiducial model of $q_{\mathrm{FIR}}(z)$ evolution incorporating the redshift evolution of cosmic-ray losses described in Section~4 and shown in Figure~\ref{fig:fig3}. We add the computed $\Delta q_{\mathrm{FIR}}(z,L_{1.4})$ in Equation~8.
    \item Model 5: Same as Model 4, but we incorporate a redshift-dependent scatter correction, approximated by using the median evolutionary parameters derived from Model 4 and using the method described in Section~5 to evaluate the correction over a grid of redshift points. This is a reasonable approximation as a correction on average of ${\sim}0.08\,\mathrm{dex}$ across all redshifts does not significantly change the luminosity/density evolution. We use a smoothing spline to interpolate the correction. The scatter correction for the PLE and LADE models are shown in Figure~\ref{fig:fig6}.
\end{enumerate}
All of these models have the same number of parameters as we have chosen to fix the intrinsic scatter $\sigma_q$ and the specific $q_{\mathrm{FIR}}(z)$ evolution. 

\begin{figure}[t!]
    \centering
    \includegraphics[width=\linewidth]{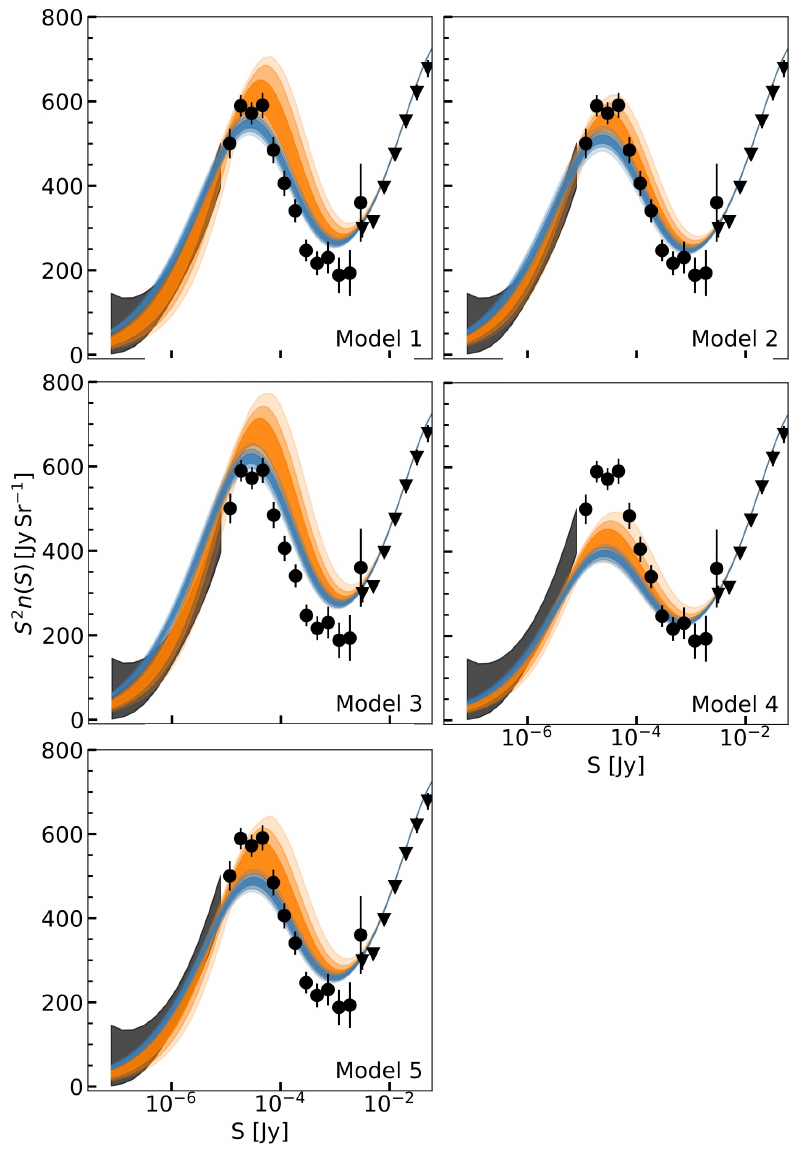}
    \caption{While $\sigma_q{\sim}0.3\,\mathrm{dex}$ is sufficient to reproduce the source count peak under both a constant $q_{\mathrm{FIR}}$ (Model 1) and one that evolves with cosmic-ray-losses (Model 5), a strictly decreasing $q_{\mathrm{FIR}}(z){\sim}(1+z)^{-0.023}$ (Model 3) substantially overestimates source counts. Without intrinsic scatter, however, a decreasing $q_{\mathrm{FIR}}(z)$ is more favored (Model 2) than a $q_{\mathrm{FIR}}(z)$ that breaks down with redshift (Model 4). The same data and color scheme is used as Figure \ref{fig:fig2}.}
    \label{fig:fig7}
\end{figure}

In Figure~\ref{fig:fig7}, we show a comparison between all 5 methods with respect to the observed DEEP2 radio source counts and the P(D) extrapolation. We find that either an intrinsic scatter correction (Model 1) to the SFRD or a decreasing $q_{\mathrm{FIR}}(z)$ out to intermediate redshifts $z{\sim}2$ without scatter (Model 2) alone can qualitatively reproduce the observed $S^2n(S)$ peak, particularly under the LADE model, while the PLE model can still underpredict source counts at the peak. However, when a scatter correction is added to Model 2 (Model 3), both the PLE and LADE models tend to overestimate source counts significantly, including near ${\sim}100\,\mathrm{\mu Jy}$. Without scatter, our fiducial model arising from redshift-evolving cosmic-ray loss effects (Model 4) significantly underpredicts source counts. Although an $L_{\star}$ galaxy will grow in luminosity with redshift and thus be less effected by non-synchrotron cosmic-ray losses, Figure \ref{fig:fig3} indicates that even $10L_{\star}$ SFGs observe a noticeable $\Delta q_{\mathrm{FIR}}>0$. Therefore, as the SFR per unit radio luminosity increases due to $\Delta q_{\mathrm{FIR}}>0$, this requires a less significant density and luminosity evolution in the local radio LF to match the observed UV/IR SFRD, resulting in a more significant underestimation of source counts compared to a nonevolving radio-FIR correlation. Such underestimation can only be resolved for the LADE model if a sufficiently large correction arising from an intrinsic scatter of $\sigma_q\approx 0.3\,\mathrm{dex}$ is incorporated (Model 5).

\begin{deluxetable*}{lcccccc}[htp!]
\tablecaption{WAIC Comparisons Relative to the Fiducial Nonevolving Model ($\mathrm{WAIC_0}$).\label{tab:waic}}
\tablehead{
\colhead{} & 
\colhead{$\mathrm{WAIC_0}$} & 
\colhead{$\Delta\mathrm{WAIC_{01}}$} & 
\colhead{$\Delta\mathrm{WAIC_{02}}$} & 
\colhead{$\Delta\mathrm{WAIC_{03}}$} & 
\colhead{$\Delta\mathrm{WAIC_{04}}$} & 
\colhead{$\Delta\mathrm{WAIC_{05}}$}
}
\startdata
PLE & $524 \pm 50$ & $-57 \pm 49$ & $-60 \pm 33$ & $10 \pm 61$ & $110 \pm 47$ & $-37 \pm 26$ \\
LADE & $519 \pm 47$ & $397 \pm 294$ & $-47 \pm 44$ & $678 \pm 304$ & $135 \pm 63$ & $122 \pm 88$ \\
\enddata
\tablecomments{We report $\Delta\mathrm{WAIC_{0i}}$ with respect to the fiducial model for models $\mathrm{i}=1,\dots,5$, with negative values indicating a better model.}
\end{deluxetable*}

\section{Discussion}

\subsection{Model Comparison}
For direct model comparison, we use the Watanabe-Akaike Information Criterion (WAIC), evaluating the likelihood of the posterior samples to the fits of the observed UV/IR SFRD history to the observed direct and P(D)-extrapolated radio source counts. It is given by
\begin{equation}
    \mathrm{WAIC} = -2\left(\sum_i \log \frac{1}{N}\sum_ np(y_i|\theta_n) -\sum_i\mathrm{V}[\log p(y_i|\theta)]\right),
\end{equation}
where the first term is the sum over the sample-averaged point-wise likelihood, and the second term is the sum of the sample variance of the point-wise likelihood, a penalty term akin to the parameter penalty term found in the ordinary Akaike Information Criterion (AIC) or Bayesian information criterion (BIC). Here, $y_i$ corresponds to the $S^2n(S)$ prediction at a given flux density $S_i$ and $\theta_n$ corresponds to a posterior parameter sample on the luminosity/density evolution. By nature of the selection of the flux density bins used to construct the radio source counts and confusion and instrumental noise, nearby source counts are likely to be correlated, but this covariance is generally unknown. We therefore assume independent $S^2n(S)$ measurements in each bin and thus assume an independent Gaussian log-likelihood. We evaluate over the grid of DEEP2 and NVSS flux bins along with points in the P(D)-extrapolated region down to $\log(S)\approx-7.1$ in steps of $\Delta \log(S)=0.2\,\mathrm{dex}$, consistent with the spacing chosen by \citetalias{Matthews_2021b}. We perform evaluations using $N=5000$ samples of the posterior. In Table \ref{tab:waic}, we show the $\Delta \mathrm{WAIC}$ with respect to the fiducial nonevolving model for each of the four models and priors along with the associated standard error in the difference. We confirm that the WAIC differences are consistent with Pareto-smoothed importance sampling leave-one-out cross validation.

We also consider a numerical comparison to the observed $q_{\mathrm{FIR}}(M_{\star},z)$ evolution (e.g., as shown in Figure \ref{fig:fig3.5}). This is because although one model may have better agreement with the radio source counts, it may not have good agreement with the observed radio-FIR correlation. We consider a comparison for a nonevolving $q_{\mathrm{FIR}}(z)$ at fixed radio luminosity (as done in Model 1), a decreasing $q_{\mathrm{FIR}}(z)$ (as done in Models 2 and 3), and our fiducial model of $q_{\mathrm{FIR}}(z)$ (as done in Models 4 and 5). As we do not have posterior samples associated with our models, we compute a simple Gaussian log-likelihood between the observations and the model-predicted $q_{\mathrm{FIR}}(z)$. This is akin to computing the Akaike Information Criterion (AIC) but without consideration of the number of free parameters, as we have fixed our model parameters. For the star formation main sequence evolution needed to convert SFRs to an equivalent stellar mass, we consider the parameterizations provided by \cite{2014ApJS..214...15S, 2015AA...575A..74S, 2024AA...691A.164K}, ensuring a conversion to a Salpeter IMF for accurate comparison (i.e., a factor $1/0.63$ for Chabrier and $1/0.67$ for Kroupa). In Table \ref{tab:lnlike} we show the resulting log-likelihood values for the various radio-FIR correlation models.

\begin{deluxetable}{lrrr}[tp!]
\tablecaption{Gaussian Log-Likelihoods for the $q_{\mathrm{FIR}}(z)$ Evolution Models
\label{tab:lnlike}}
\tablehead{
\colhead{Sample} &
\colhead{Fiducial} &
\colhead{Nonevolving} &
\colhead{Decreasing}
}
\startdata
K24 & $-250$ & $-325$ & $-389$ \\
S15 & $-269$ & $-337$ & $-401$ \\
S14 & $-299$ & $-356$ & $-420$
\enddata
\tablecomments{Larger, less-negative log-likelihoods indicate a better
fit. All models have the same number of parameters. We compare to the observed $q_{\mathrm{FIR}}(M_{\star},z)$ by \cite{2021AA...647A.123D} but shifted down by $0.3\,\mathrm{dex}$ for FIR luminosities. These values are approximate as we evaluate using the central mass in each mass bin provided by \cite{2021AA...647A.123D}.}
\tablerefs{
K24: \cite{2024AA...691A.164K};
S15: \cite{2015AA...575A..74S};
S14: \cite{2014ApJS..214...15S}
}
\end{deluxetable}

Despite a decreasing $q_{\mathrm{FIR}}(z)$ at fixed radio luminosity with minimal scatter correction being preferred for agreement with the radio source counts at a level $\gtrsim 1\sigma$ across both the PLE and LADE models (Model 2), it produces the greatest discrepancy with the observed radio-FIR correlation at fixed stellar mass as indicated by the more negative log-likelihoods. Alternatively, a radio-FIR correlation that breaks down at fixed radio luminosity provides the best agreement with observed the stellar mass-dependent radio-FIR correlation. In this case, agreement with the radio source counts at the peak requires a scatter correction larger than that implied by the observed scatter in the local radio-luminosity-dependent radio-FIR correlation \citep[e.g., $\sigma_q\approx 0.3\,\mathrm{dex}$ relative to an observed $\sigma_q\approx 0.15\,\mathrm{dex}$ at $z=0$ observed by][]{2021MNRAS.504..118M}.

This scatter need not remain constant, and could grow from ${\sim}0.15\,\mathrm{dex}$ locally to ${\sim}0.3\,\mathrm{dex}$ by $z{\sim}1$-$2$ to prevent overestimation of counts at higher flux densities ${\sim}100\,\mathrm{\mu Jy}$ contributed by lower redshift radio LFs and match the source count peak at lower flux densities ${\sim}30\,\mathrm{\mu Jy}$ contributed by higher redshift radio LFs. This increase may reflect increasing variations in ISM conditions (including magnetic field strengths, ISM densities, and scale heights) that lead to larger variations in the synchrotron suppression at higher redshifts and spectral indices required for $k$-corrections. This is also in contrast to \cite{2021AA...647A.123D} who infer that, at a fixed stellar mass, the radio-FIR correlation appears to maintain a ${\sim}0.2\,\mathrm{dex}$ scatter with redshift.

Due to the large variance in the posterior samples driving up the absolute WAIC statistic and thus its standard error, no model produces a statistically significant improvement at a level $>3\sigma$ in the agreement with the radio source counts with respect to the nonevolving radio-FIR correlation without an intrinsic scatter correction.

\subsection{Impact of Radio LF Evolution}

Naturally, the choice of the radio LF evolution matters significantly, for the PLE model appears to be less favored relative to the LADE, particularly for a $q_{\mathrm{FIR}}(z)$ that breaks down with redshift; even with a $\sigma_q=0.3\,\mathrm{dex}$ scatter correction, the PLE model is unable to recover the source count peak. The inconsistency of a PLE model is also supported by \cite{2024A&A...683A.174W} from direct fits to 3 GHz LFs in the VLA-COSMOS field.

We thus consider the opposite extreme, where the density evolution is significantly stronger, causing the luminosity evolution to be stronger and thus produce more sources at higher flux density at lower redshifts. Rather than incorporating no prior on the density and luminosity evolution, we now enforce a prior that $g(z)$ drops by $0.3,0.2,0.1\times$ by $z=3$, generally consistent with the 150 MHz LFs observed by \cite{10.1093/mnras/stad1602} and the 3 GHz LFs fitted by \cite{2024A&A...683A.174W}. For simplicity, we consider the three prescribed $q_{\mathrm{FIR}}(z)$ models without a scatter correction. Again, as with comparisons of intrinsic scatter, comparisons to the $L_{\star}$ evolution between 150 MHz and 1.4 GHz LFs should be interpreted with caution as the SFG samples and shapes of the radio LFs may be different due to different radio emission being traced. However, under the assumption of a typical $\alpha=-0.7$ low frequency spectral index \citep[e.g.,][]{2017MNRAS.469.3468C}, a conversion in the $L_{\star}$ evolution at 150 MHz and 1.4 GHz can be made \citep[e.g., as done in][but for 3 GHz LFs]{2024A&A...683A.174W}, as variations arising to the fitted model (LADE vs. PLE) rather than the spectral index assumption dominate the inferred $L_{\star}$ evolution.

\begin{figure}[tp!]
    \centering
    \includegraphics[width=\linewidth]{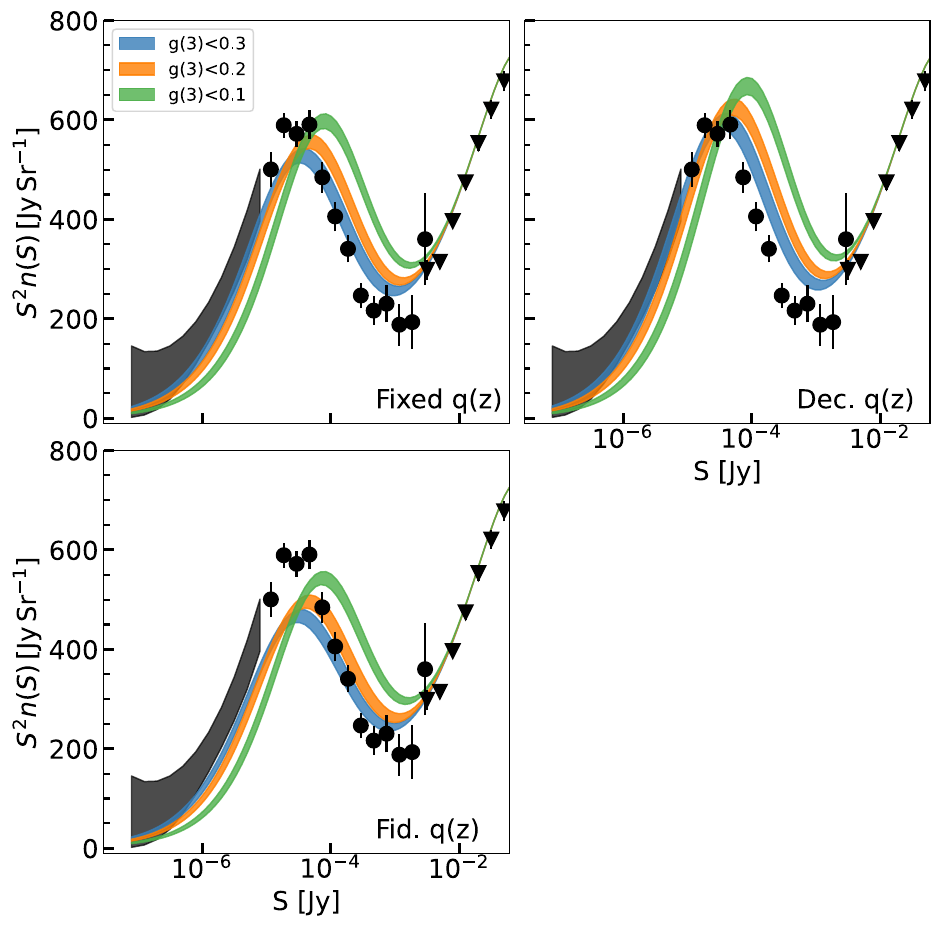}
    \caption{Source count comparisons for a fixed, decreasing, and our fiducial $q_{\mathrm{FIR}}(z)$ model when imposing a strong density evolution determined by the upper limit on $g(z=3)$ (differentiated by color). As the density evolution strengthens, the source count peak shifts to lower flux densities across all models and is thus disfavored. For ease of comparison, we have only plotted the $1\sigma$ posterior intervals. The observed source count data (black) are the same as that in Figure \ref{fig:fig2}.}
    \label{fig:fig7.5}
\end{figure}

The resulting comparison is shown in Figure \ref{fig:fig7.5}, indicating that none of the prescribed $q_{\mathrm{FIR}}(z)$ evolutionary models support a strong density evolution with $g(z=3)\lesssim 0.1$. Additionally, a constant scatter correction is not sufficient to correct for such a discrepancy as it will only increase the source counts at higher flux densities $\gtrsim 100\,\mathrm{\mu Jy}$, despite already overestimating them by $>3\sigma$, particularly for the fixed and decreasing $q_{\mathrm{FIR}}(z)$. We observe that even in the limit of a stronger density evolution with $0.2\lesssim g(z=3)\lesssim 0.3$, our fiducial $q_{\mathrm{FIR}}(z)$ model requires an intrinsic scatter $\sigma_q\approx 0.3\,\mathrm{dex}$ to account for the source count peak, resulting in a ${\sim}20$--$25\%$ increase in the source count contribution. However, as discussed in the previous section, to avoid overestimating source counts near ${\sim}100\,\mathrm{\mu Jy}$ contributed by sources at lower redshifts, the intrinsic scatter must grow with redshift from $z=0$.

\subsection{Evaluating the Effect of $ q_0(L_{1.4},z)$}
As discussed previously, our model, similar to \cite{2009ApJ...706..482M, 2024ApJ...975...15Y}, ignores the possibility of a changing calibration between FIR and thermal radio emission $q_0$ at fixed total radio luminosity. Under this assumption, the change in thermal emission directly tracks the change in $q_{\mathrm{FIR}}$. In this section, we discuss the effect of this assumption on the fiducial radio-FIR correlation evolution.

Locally, our prescribed $f_t(L_{1.4}, z=0)$ relation is steeper than $q_{\mathrm{FIR}}(L_{1.4},z=0)$. This implies that for $L\lesssim L_{\star}$, IR emission underestimates SFRs relative to thermal radio emission. Alternatively, at $L\gtrsim L_{\star}$, thermal radio emission may underestimate SFRs relative to IR emission or there must be an excess of IR emission per unit SFR. This also appears to be true for local luminous starbursts, where for example, SFGs $L_{1.4}\geq 10^{22.5}\,\mathrm{W\, Hz^{-1}}$ in the sample of \cite{2018MNRAS.474..779G} exhibit typical thermal fractions of ${\sim}4\%$ (with some as low as ${\sim}1\%$), which under a calibration $q_0\approx 3.32$ would imply $q_{\mathrm{FIR}}\approx 1.92$, which is below the \citetalias{Matthews_2021b} value of $q_{\mathrm{FIR}}\approx 2.19$ and even lower than the average $q_{\mathrm{FIR}}\approx 2.26$ among that sample. Hence, $q_0$ may be larger at higher radio luminosities in order to match the observed radio-FIR correlation for the observed thermal fractions. We consider how $q_0$ could change with redshift in these different luminosity regimes.

\subsubsection{Effect at Low Radio Luminosities}

At the low luminosity end, the smaller $q_0<3.32$ can be attributed to a reduced dust content resulting in the reduced reprocessing efficiency of UV emission \citep[e.g.,][]{2003ApJ...586..794B, 2010ApJ...717..196L}. For example, Equation~14 implies $f_t(L_{1.4}=10^{20}\,\mathrm{W\, Hz^{-1}}, z=0)\approx 0.26$, which corresponds to $\Delta\log(f_t)\approx 0.37\,\mathrm{dex}$ relative to $L_{\star}(z=0)$. This is compared to a $\Delta q_{\mathrm{FIR}}=0.17\,\mathrm{dex}$ relative to $L_{\star}$, suggesting that IR emission underestimates SFRs by ${\sim}0.2\,\mathrm{dex}$ relative to thermal radio emission based on the $L_{\star}$ calibration, requiring $q_0\approx 3.12$ instead of $q_0=3.32$. This underestimation is generally consistent with that implied by the IRX--$M_{\star}$ relation, assuming that most low luminosity SFGs lie on the local star forming main sequence. By solving the implicit equation
\begin{equation}
    \mathrm{SFR_{IR}}(L_{1.4})/f_{\mathrm{TIR}}(M_{\star}) = \mathrm{SFR_{\mathrm{MS}}}(M_{\star}),
\end{equation}
we find that at $z=0$, $L_{1.4}=10^{20}\,\mathrm{W\, Hz^{-1}}$ corresponds to $\log(M_{\star}/M_{\odot})\approx 8.9$ while $L_{1.4}=L_{\star}$ corresponds to $\log(M_{\star}/M_{\odot})\approx 9.7$. We have assumed the \cite{2015AA...575A..74S} main sequence relation, so these masses are on the Salpeter IMF. This corresponds to a $\Delta \log(f_{\mathrm{TIR}})\approx 0.19\,\mathrm{dex}$, generally consistent with the thermal fraction change relative to $L_{\star}$. We note that this result may vary by a few hundredths of a dex due to different star forming main sequence and thermal fraction relations and their uncertainties. Assuming that thermal radio emission is a good SFR tracer in this luminosity regime, we therefore take $\Delta\log(f_t)\approx \Delta\log(\mathrm{SFR_{MS}})$, enabling us to infer the change in stellar mass producing the same radio luminosity out to redshift $z$. We can then apply the IRX--$M_{\star}$ relation to estimate the extra factor by which IR emission underestimates SFRs relative to thermal radio emission at redshift $z$ because the typical stellar mass producing the same radio emission has dropped.

Therefore, for $L_{1.4}=10^{20}\,\mathrm{W\, Hz^{-1}}$, by $z=3$, $\Delta \log(f_t)\approx \Delta\mathrm{SFR}_{\mathrm{MS}}\approx 0.36\,\mathrm{dex}$. If this SFG starts at $\log(M_{\star}/M_{\odot})\approx 8.9$, then the SFR main sequence evolution of \cite{2015AA...575A..74S} implies $\Delta \log(M_{\star}/M_{\odot})\approx -0.54\,\mathrm{dex}$. From Equations~20 and 21, which are assumed to be approximately fixed with redshift, this implies $\Delta \log (f_{\mathrm{TIR}}) \approx \Delta q_0\approx -0.22\,\mathrm{dex}$. Alternatively, for $L_{1.4}\approx 2\times10^{21}\,\mathrm{W\, Hz^{-1}},$ which corresponds to $\log(M_{\star}/M_{\odot})\approx 9.8$ at $z=0$, a $\Delta \log(f_t)=0.33\,\mathrm{dex}$ implies $\Delta \log(M_{\star}/M_{\odot})\approx -0.61\,\mathrm{dex}$, but because the IRX is more saturated at higher stellar masses, this corresponds to only $\Delta \log(f_{\mathrm{TIR}})\approx \Delta q_0\approx -0.12\,\mathrm{dex}$, so the correction is smaller. Thus, for galaxies where thermal radio emission is expected to be a good SFR tracer, $q_0$, while expected to decrease with redshift, should see a weaker decrease with increasing total radio luminosity. In any case, the $\Delta q_{\mathrm{FIR}}$ driven by synchrotron suppression is larger than that driven by changes in the relative efficiency of IR and thermal radio emission in tracing SFRs. A ``breakdown'' is still required at fixed radio luminosity for $L_{1.4}\lesssim L_{\star}(z=0)$ in order to ensure agreement with $q_{\mathrm{FIR}}(M_{\star}, z)$ at low stellar masses.

The physical limit on thermal fractions at $100\%$ has implications on $q_{\mathrm{FIR}}$ in low luminosity dwarf SFGs. If $q_{\mathrm{FIR}}$ continues to rise with lower radio luminosities for $L_{1.4}\ll 10^{20}\,\mathrm{W\, Hz^{-1}}$ while the growth in thermal fractions start to flatten, then at some point the assumed $q_0\approx 3.32$ may hold again. This could arise due to increased Lyman continuum escape from HII regions before ionizing any neutral hydrogen, particularly in metal-poor dwarf SFGs \citep[e.g.,][]{2022A&A...667A..35R} that would similarly reduce the effectiveness of thermal radio emission as an SFR tracer. Alternatively, $q_{\mathrm{FIR}}$ may not be significantly higher compared to normal SFGs, in which case $q_0$ would saturate at low radio luminosities at a value lower than the assumed $q_0=3.32$ at $L_{\star}.$

\subsubsection{Effect at High Radio Luminosities}
For more luminous SFGs, (e.g., $L_{1.4}\gtrsim 7\times10^{21}\,\mathrm{W\, Hz^{-1}}$), the IR SFRs imply that they are not on the $z=0$ SFG main sequence of \cite{2015AA...575A..74S} via Equation~26 and are thus likely dominated by dusty starbursts or (ultra)luminous infrared galaxies (LIRGs). If thermal fractions continue to decrease at these higher radio luminosities \citep[e.g., as suggested by][]{2018MNRAS.474..779G}, then the local radio-FIR correlation implies that thermal radio emission becomes a poorer tracer of SFR for increasingly luminous SFGs relative to FIR emission and the assumed calibration. Therefore, $\Delta q_0$ in this luminosity regime may instead be driven by how the fraction of ionizing photons absorbed by dust before ionizing neutral hydrogen changes \citep[e.g.,][]{Inoue_2001, 2003ApJ...583..727D}, which would impact the amount of thermal radio emission per unit SFR. Alternatively, there may also be an excess of FIR emission not from star formation (e.g., radio quiet AGN or dust heating by old stars).

In terms of Lyman continuum absorption by dust, from \cite{Inoue_2001}, the optical depth of dust to Lyman continuum photons in an HII region is $\tau_d\propto \mathcal{D} N_H$, where $\mathcal{D}$ is the dust-to-gas mass ratio, which is inferred to be related to the $\log(\mathrm{O/H})$ abundance, and $N_H$ is the hydrogen column through an HII region. While the dust-to-gas mass ratio may decrease given the lower metallicities at higher redshifts (e.g., \citetalias{2014ARA&A..52..415M}) and as SFGs gradually move down the star forming main sequence at fixed radio luminosity, the hydrogen column will grow due to the increasing HII region densities \citep[e.g.,][]{Isobe_2023}, which provide two competing effects on the optical depth with redshift and thus the fraction of dust-absorbed Lyman continuum photons. While this tradeoff can explain why, for example, any redshift evolution in the IRX-$M_{\star}$ relation is mild for typical SFGs, it is not well known how this would change for SFGs well above the star forming main sequence.

It is also difficult to infer any evolution in the contamination by older stellar populations to the FIR emission. Locally, more optically-luminous SFGs are more contaminated by older stellar populations (with large scatter). On the other hand, there appears to be minimal correlation between the contamination fraction and TIR luminosity and hence total radio emission \cite{2003ApJ...586..794B}, so it is unlikely that the increase in $q_0$ locally is driven by an increasing contamination by older stellar populations.

AGN can appear in more luminous SFGs such as ULIRGs, but they typically contribute $\lesssim 30\%$ of the bolometric flux and are generally subdominant to dust emission in the FIR even when radio loud \citep{Dey_2024}. SFGs selected based on $q_{\mathrm{FIR}}$ cuts \citep[e.g., as done by][and thus used for the derivation of Equation 8]{2019ApJ...872..148C} will not contain radio-loud AGN by definition \citep{2026A&A...706A.111P}, but it is unknown what fraction of these SFGs in large samples can still exhibit an AGN FIR excess.

Consequently, $\Delta q_0$ above $L_{\star}(z=0)$ is not easily determined and requires further work. Requiring agreement with the stellar-mass dependent radio-FIR correlation out to $z{\sim}3$ at the high mass end, however, would seem to suggest that these competing effects prefer a slightly negative $\Delta q_0 \gtrsim -0.1\,\mathrm{dex}$ for SFGs near $L_{1.4}{\sim}10^{24}\,\mathrm{W\, Hz^{-1}}$ by $z\approx 3$. In this case, while optical depths to dust or contamination by older stars may slightly decrease with redshift, this may also be attributed to such SFGs gradually falling onto the star forming main sequence by $z\approx 3$ \citep[e.g.,][]{2023A&A...679A..47S, Tabatabaei_2025}, causing FIR emission to be a gradually less optimal SFR tracer. This could also be explained by a weaker or even decreasing thermal fraction with redshift, which would require magnetic field strengths to grow more rapidly relative to ISM densities than assumed in the model to ensure that synchrotron losses remain dominant.

%\textcolor{red}{SFGs with radio luminosities following along $L_{\star}(z) = L_{\star}(z=0)f(z)$ will fall down onto the star forming main sequence at higher redshifts, with $L_{\star}(z=3)$ corresponding to similar stellar masses on the star forming main sequence as $L_{\star}(z=0)$ as inferred by their IR SFRs under our fiducial model (thus being self-consistent). Therefore, we do not expect a significant change in $q_0$ along the $L_{\star}$ track due to changes in the efficiency of IR emission as a star formation tracer, assuming that the IRX-$M_{\star}$ relation does not change significantly with redshift. If stellar mass is a proxy for optical depths to dust on the star forming main sequence then we also do not expect significant change in $q_0$ due to changes in the efficiency of thermal radio emission due to Lyman continuum absorption by dust.}

Additionally, under our fiducial radio-FIR correlation evolution and the LADE prior, we can estimate the contribution of SFGs corresponding to SFRs that cannot be uniquely placed on the star formation main sequence to the total radio luminosity density to gauge how much an unknown $\Delta q_0$ could affect the SFRD evolution. This can equivalently be thought of as the contribution by luminous (nondwarf) starbursts and ULIRGs. At $z=0, 1$ and $2$, we find upper radio luminosity limits of $L_{1.4}\approx 7.4\times10^{21}, 1.5\times10^{23}, 7.3\times10^{23}\,\mathrm{W\, Hz^{-1}},$ respectively. Based on the LADE evolution, this corresponds to ${\sim} 53\%, 39\%, 16\%$ of the total integrated luminosity density at each redshift, respectively. Assuming that $\Delta q_{\mathrm{FIR}}(z)$ is milder at high radio luminosities, we do not expect uncertainties on its evolution among high radio luminosity SFGs at high redshifts $z\gtrsim 2$ to strongly impact the cosmic star formation history, especially as the luminosity evolution is expected to turn over.

In summary, for main sequence SFGs below or near $L_{\star}(z=0)$, the evolving SFR main sequence and IRX--$M_{\star}$ relation implies downward corrections of ${\sim}0.1$-$0.2\,\mathrm{dex}$ to our derived $\Delta q_{\mathrm{FIR}}(z)$ by $z=3$. These corrections, however, are still smaller than the predicted increase in $\log(f_t)$ so the net $\Delta q_{\mathrm{FIR}}$ remains positive by $z=3$ but is somewhat dampened. 
At higher radio luminosities, where dust obscuration is nearly saturated, evolution in $q_0$ at fixed radio luminosity may be more sensitive to the amount of Lyman continuum photons absorbed by dust or excess FIR emission from dust heating by old stars or AGN whose redshift evolution is not well understood. In this limit, the sign of $\Delta q_0$ cannot be determined robustly, but its magnitude is likely to be small given the weak SFR evolution and general agreement with the stellar mass-dependent radio-FIR correlation.

\section{Conclusion}
We have provided several modifications to the original luminosity and density evolution of the local SFG radio LF constrained by radio source counts in \citetalias{Matthews_2021b}, which appear to indicate a stronger evolution in the cosmic SFRD for redshifts below $z{\sim}2$. We show that a prior-free PLE and LADE model consistent with the UV/IR--derived SFRD from a compilation of measurements from \citetalias{2014ARA&A..52..415M} and more recent IR surveys can underestimate source counts near the $S{\sim}30\,\mathrm{\mu Jy}$ peak in $S^2n(S)$ by $\gtrsim 3\sigma$.

To examine the nature of the source count discrepancy and what it can tell us about the radio-FIR correlation, we introduce various models of a redshift-evolving radio-FIR correlation that is parameterized purely in terms of redshift and total radio luminosity at 1.4 GHz, two directly observable quantities. We consider a weakly decreasing $q_{\mathrm{FIR}}(z)$ at fixed radio luminosity consistent with \cite{2021AA...647A.123D}, a nonevolving $q_{\mathrm{FIR}}(z)$, and a $q_{\mathrm{FIR}}(z)$ that breaks down due to increases in the radio thermal fraction with redshift. These changes arise from redshift-dependent cosmic-ray loss effects including ionization, bremsstrahlung, escape, synchrotron, and inverse Compton, which generally lead to $q_{\mathrm{FIR}}(z)$ increasing with redshift at \textit{fixed radio luminosity.}

While we have assumed that the FIR-to-thermal radio emission calibration $q_0$ does not significantly change at fixed radio luminosity, more sophisticated models of the radio-FIR correlation might incorporate such possibility. FIR emission from less luminous SFGs $L_{1.4}\ll L_{\star}(z=0)$ underestimates SFRs \citep[e.g.,][]{2003ApJ...586..794B, 10.1093/mnras/stu896} due to a significant portion of star formation being dust unobscured. Hence, evolution towards lower stellar masses producing the same radio luminosity will result in an increase in the underestimation of SFR by IR emission alone, causing $q_0$ to decrease slightly with redshift. For $L_{1.4}\gg L_{\star}(z=0)$, any evolution in $q_0$ at fixed radio luminosity is likely weak but difficult to determine due to the competing effects of decreasing metallicities but increasing gas densities with redshift. Additionally, changes in excess FIR emission from AGN contamination or dust heating by older stellar populations is not well known. These effects complicate the use of the radio-FIR correlation in converting radio LFs into SFRDs due to the requirement of a radio-luminosity dependent radio-FIR correlation evolution.

For each evolution, we also consider the effect of intrinsic scatter in the radio-FIR correlation, which is expected to artificially boost SFRDs when derived from radio LFs due to the Eddington bias. Our main conclusions are:

\begin{enumerate}
    \item When the radio LF evolution is set by the observed UV/IR SFRD, radio source counts are sensitive to the prescribed $q_{\mathrm{FIR}}(z)$ evolution before consideration of any intrinsic scatter. While a decreasing $q_{\mathrm{FIR}}(z)\propto (1+z)^{-0.023}$ brings source counts in agreement to within $1\sigma$ at the $S^2n(S)$ peak, an increasing $q_{\mathrm{FIR}}(z)$ consistent with our cosmic-ray loss model produces a ${\sim}4$-$5\sigma$ discrepancy with the peak source counts.

    \item While a decreasing $q_{\mathrm{FIR}}(z)$ without scatter produces the best agreement with the radio source counts for both the PLE and LADE models, showing a $\Delta \mathrm{WAIC}<0$ with respect to the fiducial nonevolving model, direct comparisons to the observed radio-FIR correlation from \cite{2021AA...647A.123D} at fixed stellar mass prefer a radio-FIR correlation that breaks down with redshift (with $\Delta \log(L)>50$). A decreasing $q_{\mathrm{FIR}}(z)$ at fixed stellar mass does not necessarily require a decreasing $q_{\mathrm{FIR}}(z)$ because of the local nonlinear radio-FIR correlation. As the SFR per stellar mass rises rapidly with redshift, SFGs at fixed mass on the main sequence will generally produce more radio emission and thus migrate to lower $q_{\mathrm{FIR}}$ initially.
    
    \item Irrespective of the radio LF evolution prior, our fiducial $q_{\mathrm{FIR}}(z)$ model remains $>3\sigma$ discrepant with the $S^2n(S)$ source count peak, requiring an intrinsic scatter $\sigma_q\approx 0.3\,\mathrm{dex}$ at $z\gtrsim 1$ to reproduce it. This is higher than the observed local $\sigma_q\approx 0.15\,\mathrm{dex}$ found by \cite{2021MNRAS.504..118M} on the nonlinear radio-FIR correlation, and may suggest an intrinsic scatter that is higher at higher redshifts. This would reflect a broader distribution of ISM conditions, including magnetic field strengths, ISM densities, and disk sizes, that would all contribute to a broader distribution of thermal radio fraction or spectral indices.

    \item Enforcing agreement between the radio source counts and the UV/IR SFRD disfavors a strong density evolution, requiring that $g(z=3)\gtrsim 0.1$. Irrespective of the applied $q_{\mathrm{FIR}}(z)$ evolution and intrinsic scatter, the source count peak is predicted to shift to higher flux densities. This suggests the possibility of additionally using precise radio source counts to break the degeneracy in the radio LF evolution.

    \item More work needs to be done to understand the origin of the changes in the FIR-to-thermal radio calibration $q_0$ that is assumed for $L_{\star}(z=0)$ galaxies both locally as a function of total radio luminosity and with redshift at fixed radio luminosity. This can affect how $\Delta q_{\mathrm{FIR}}(z)$ evolves on top of changes in the thermal fraction with redshift. These steps are ultimately required to move towards understanding and using radio emission alone as a cosmic star formation tracer independent of IR emission.
\end{enumerate}

Ultimately, we have shown that compilations of a large number of UV/IR SFRD measurements and comparisons to radio source counts can provide further insight into the discrepancies between converting radio and IR emission into SFRs, particularly for $L_{\star}$ galaxies. However, degeneracies remain between the amount of intrinsic scatter, the prescribed $q_{\mathrm{FIR}}(z)$ evolution, and the specific radio LF evolution, requiring additional constraints to break such degeneracies. Deeper radio source counts observed with upcoming radio telescopes like the DSA, ngVLA, and SKA--either thermal noise-limited or confusion noise-limited aided by a P(D) analysis--may be able to place stronger constraints on the allowed evolutionary models of the radio LF out to higher redshifts. As shown in Appendix A, the P(D)-extrapolated source counts $\ll 10\,\mathrm{\mu Jy}$ receive significant contributions from LFs at $z\gtrsim 2$. Specific constraints on the P(D) distribution and hence the allowed radio LF evolution of SFGs is subject to future work, requiring forward modeling of the deep fields \citep[e.g., as done in][]{Matthews_2021a}. Alternatively, higher redshift observations of fainter SFGs may place tighter constraints on their thermal fraction and thus the radio-FIR correlation evolution.

\begin{acknowledgments}
We thank the anonymous referee for providing in-depth feedback that helped improve this work significantly. This work was funded in part by Schmidt Sciences. This paper depended on source count data originally obtained from MeerKAT and the National Radio Astronomy Observatory's Karl G. Jansky Very Large Array (VLA). The MeerKAT telescope is operated by the South African Radio Astronomy Observatory, which is a facility
of the National Research Foundation, an agency of the Department of Science and Innovation. The National Radio Astronomy Observatory and Green Bank Observatory are facilities of the U.S. National Science Foundation operated under cooperative agreement by Associated Universities, Inc. We additionally thank George Helou for useful discussions that helped improve this paper.
\end{acknowledgments}

\facilities{MeerKAT, VLA}

\software{arviz \citep{Martin2026}, astropy \citep{2022ApJ...935..167A}, emcee \citep{2013PASP..125..306F}, scipy \cite{2020SciPy-NMeth}, numba \citep{2015llvm.confE...1L}}


\newpage
\appendix
\section{Source Count Contributions at Each Redshift}
\begin{figure}[htp!]
    \centering
    \includegraphics[width=\linewidth]{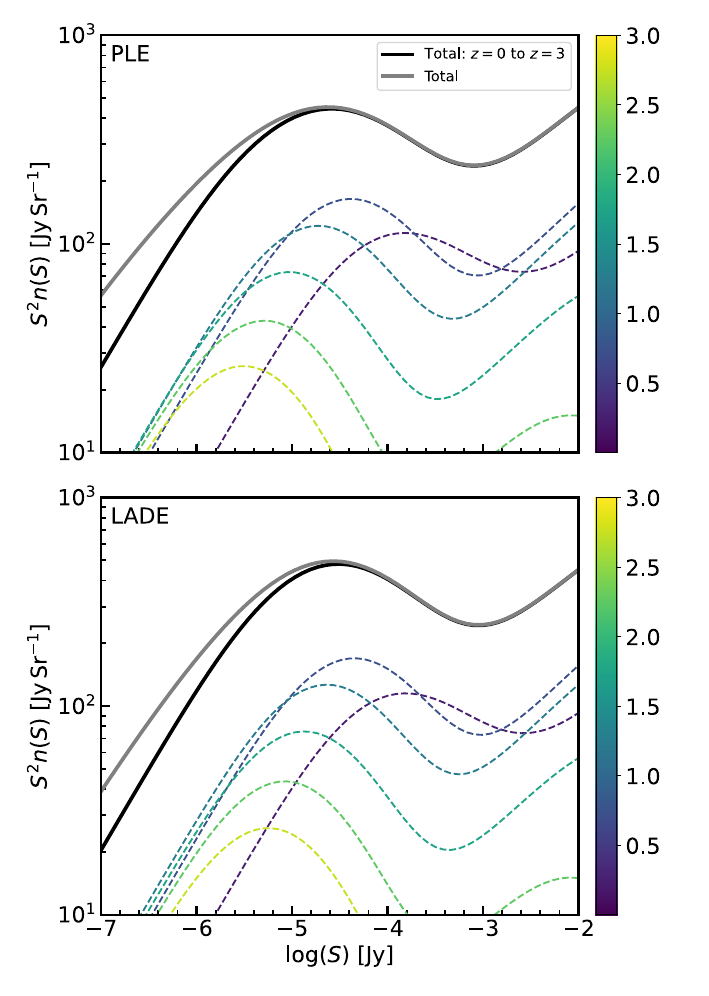}
    \caption{Contributions to the source counts under PLE (top) and LADE (bottom) in $\Delta z=0.5$ bins out to $z=3$ (colored, dashed) focusing on the SFG contribution. In both cases we compare the integrated source count out to $z=3$ (black, solid) and out to $z\to \infty$ (gray, solid), indicating that a radio LF evolution out to $z=3$ is generally sufficient to constrain source counts near the SFG peak.}
    \label{fig:fig A1}
\end{figure}

We show that a prescribed luminosity/density evolution of the local radio LF is sufficient to constrain the peak in the radio source counts observed near ${\sim}30\,\mathrm{\mu Jy}$. We adopt the median evolutionary parameters from both the PLE and LADE models and consider integrations in redshift bins of $\Delta z=0.5$ from $z=0$ to $z=3$ in both cases. In Figure \ref{fig:fig A1}, we show the individual contributions in each $\Delta z$ bin along with the summed contribution from $z=0$ to $z=3$ and the summed contribution from $z=0$ to $z\to \infty$. In both the PLE and LADE models, the radio LF evolution out to $z=3$ is sufficient to match the full source counts indicated by integrating out to $z\to \infty$ down to the edge of the direct source count limit at $\log(S)=-5$, but deviate in the P(D)-extrapolated region. Thus, a radio LF evolution constraint out to $z=3$ is generally sufficient to investigate the source count discrepancy near the peak. 

\section{Radio Thermal Fraction as a Function of Radio Luminosity}
We use the SFG sample of \cite{2017ApJ...836..185T} for which radio thermal fractions at $1.4\,\mathrm{GHz}$ were derived through direct constructions of radio SEDs up to 10 GHz and separately using H$\alpha$ luminosities, which are directly correlated to the thermal radio luminosity once dust corrected. We only fit using galaxies that were not identified to have AGN nuclear activity ($n=14$). $k$-corrections are not applied given the small distances $D\lesssim 20\,\mathrm{Mpc}$. We use the \textsc{scipy} orthogonal distance regression (ODR) routine to estimate the parameter $\beta$ in Equation 13, accounting for uncertainty in the luminosities and the thermal fractions. The resulting fit to the radio spectrum-derived thermal fractions is shown in Figure \ref{fig:fig A1} with $\beta=0.39\pm 0.07$, for which we take the central value as an approximation.

\begin{figure}[h!]
    \centering
    \includegraphics[width=\linewidth]{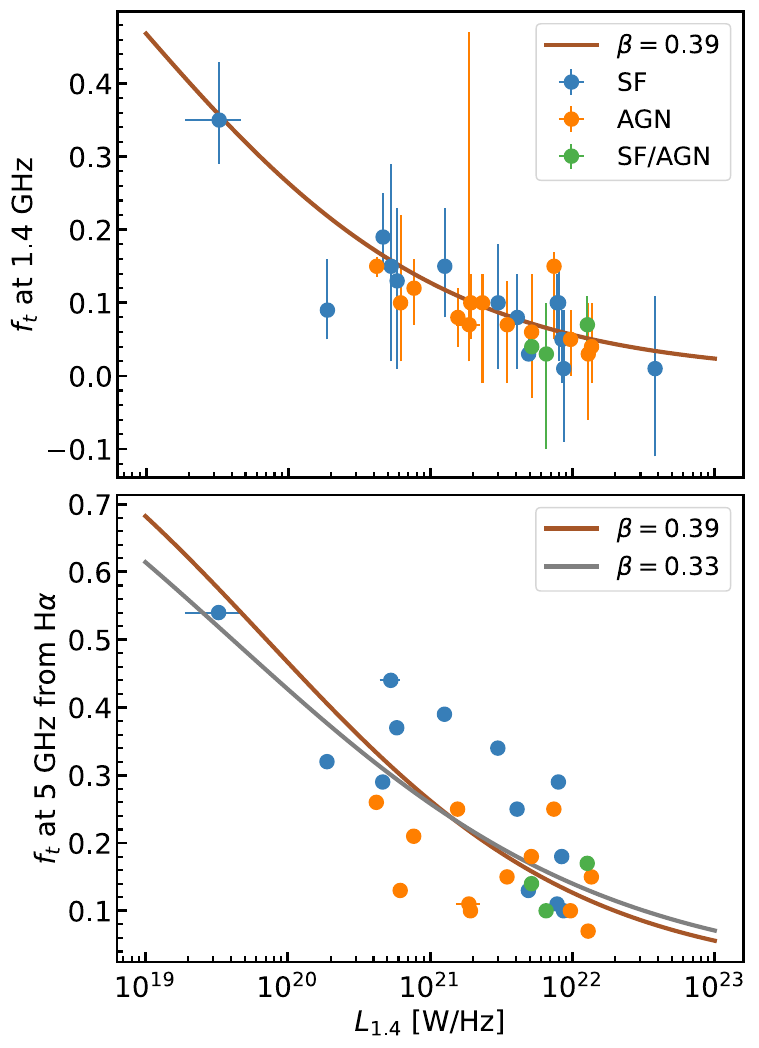}
    \caption{Change in thermal fraction as a function of 1.4 GHz spectral luminosity derived using the KINGFISHER sample among galaxies selected to have no AGN nuclear activity from \citep{2017ApJ...836..185T} using both direct fits to the radio SED (top) and implied by dust-corrected H$\alpha$ luminosities at 5 GHz (bottom); uncertainties are not originally provided. The chosen slope $\beta$ is visually consistent with both methods. We include additional points for galaxies that exhibit evidence of AGN nuclear activity, which were not used for fitting, but appear consistent with the SFGs.}
    \label{fig:fig A2}
\end{figure}

We obtain a consistent slope $\beta$ when the high thermal fraction point at low luminosity is removed ($\beta=0.40\pm 0.12$) as well as if galaxies with nuclear type ``AGN" and ``SF/AGN" are included with the SF-only galaxies yielding ($\beta=0.35\pm 0.04$). The slope $\beta$ is also consistent with that inferred by the dust-corrected H$\alpha$ luminosities at 5 GHz, for which we obtain $\beta = 0.33 \pm 0.07$, shown in the bottom panel of Figure \ref{fig:fig A2}. A better consensus of how the thermal fraction changes with radio luminosities requires a larger sample of SFGs without nuclear AGN activity.

Given that $q\propto q_0+\log(f_t)$, an increase in the thermal fraction down to lower radio luminosities could drive the stellar-mass dependence in the radio-FIR correlation. The growth in thermal emission must be stronger than the factor by which the FIR luminosity per unit SFR is reduced due to less massive SFGs being more optically thin to UV emission \citep{2003ApJ...586..794B} and thus less efficient at reprocessing UV emission into IR in order to cause $q_{\mathrm{FIR}}$ to rise with radio luminosity.

\clearpage

\bibliography{sample701}{}
\bibliographystyle{aasjournalv7}



\end{document}